
\input psfig.sty

%
%
%

\ifx\mnmacrosloaded\undefined \input mn\fi

%

\newif\ifAMStwofonts

\ifCUPmtplainloaded \else
  \NewTextAlphabet{textbfit} {cmbxti10} {}
  \NewTextAlphabet{textbfss} {cmssbx10} {}
  \NewMathAlphabet{mathbfit} {cmbxti10} {} 
  \NewMathAlphabet{mathbfss} {cmssbx10} {} 
  \ifAMStwofonts
    \NewSymbolFont{upmath} {eurm10}
    \NewSymbolFont{AMSa} {msam10}
    \NewMathSymbol{\upi}     {0}{upmath}{19}
    \NewMathSymbol{\umu}     {0}{upmath}{16}
    \NewMathSymbol{\upartial}{0}{upmath}{40}
    \NewMathSymbol{\leqslant}{3}{AMSa}{36}
    \NewMathSymbol{\geqslant}{3}{AMSa}{3E}

  \else
    \def\umu{\mu}
    \def\upi{\pi}
    \def\upartial{\partial}
  \fi
\fi


\pageoffset{-2.5pc}{0pc}

\loadboldmathnames



 \onecolumn        
\pagerange{}    
\pubyear{2001}
\volume{}

\begintopmatter  

\title{Bar strengths in spiral galaxies estimated from 2MASS images}
\author{Eija Laurikainen, Heikki Salo}
\affiliation{Division of Astronomy, Dep. of Phys. Sci, FIN-90014, Finland}

\shortauthor{E. Laurikainen, H. Salo}
\shorttitle{Bar strengths}



\abstract

Non-axisymmetric forces are presented for a sample of 107 spiral galaxies,
of which 31 are barred (SB) and 53 show nuclear activity.
As a database we use JHK images from the 2 Micron All Sky Survey, and  
the non-axisymmetries are characterized by the ratio of the tangential 
force to the mean axisymmetric radial force field, following Buta $\&$
Block (2001). Bar strengths 
have an important role in many extra-galactic problems and therefore
it is important to verify that the different numerical methods applied
for calculating the forces give mutually consistent results. We
apply both direct Cartesian integration and a polar grid integration
utilizing a limited number of azimuthal Fourier components of density. 
We find that bar strength is independent of 
the method used to evaluate the gravitational potential. 
However, because of the distance dependent smoothing by Fourier 
decomposition the polar method is more suitable for weak and noisy images. 
The largest source of uncertainty in the derived bar strength appears to 
be the uncertainty in the vertical scale-height, which is difficult to measure 
directly for most galaxies. On the other hand, the derived bar stength is
rather insensitive to the possible gradient in the vertical scale-height 
of the disk or to the exact model of the vertical density distribution, 
provided that a same effective vertical dispersion is assumed in all models.  
In comparison to the pioneering study by Buta $\&$ Block (2001),
bar strength estimate is here improved by taking 
into account the dependence of the vertical scale-height on the Hubble 
type: we find that for thin disks bar strengths are stronger than for
thick disks by an amount which may correspond to even one bar strength class.

We confirm the previous result by Buta $\&$ Block (2001) and 
Block et al. (2001) showing that the dispersion in bar strength is large
among all the de Vaucouleurs's optical bar classes.  
In the near-IR 40 $\%$ of the galaxies in our sample have bars 
(showing constant phases in the m=2 Fourier amplitudes in the bar region), 
while in the optical 1/3 of these bars are obscured by dust.
Significant non-axisymmetric forces can be induced also by the  
spiral arms, generally in the outer parts of the galactic disks, which may
have important implications on galaxy evolution.
Possible biases
of the selected sample are also studied: we find that the number of
identified bars rapidly drops when the inclination of the 
galactic disk is larger
than $50^0$. A similar bias is found in The Third Reference Catalogue of
Bright Galaxies, which might be 
of interest when comparing bar frequencies at high and low redshifts.


\maketitle  

\vfill
\eject

\section{Introduction}

Bars consist mostly of old stellar population (de Vaucouleurs 1955; 
Elmegreen $\&$ Elmegreen 1985), which stresses their significance as 
dynamically important components in galaxies. In fact, a large fraction
of galaxies have bars (Block $\&$ Wainscoat 1991; Knapen et al. 2000;
Eskridge et al. 2000, Block et al. 2001), indicating that they must be 
long-lived phenomena in galaxies. Bars are fundamental in galaxy evolution,
suggested to be driving forces for star formation, formation
of rings and global spiral density waves, and even for the onset of nuclear 
activity. When quantified the correlations between bar strength and the
other properties of the galaxies can be studied.
The wavelength that best traces the dynamical mass 
is the near-IR, where the obscuration of dust is also less 
significant than in the optical region. For example, galaxies like 
NGC 5195, which are 
irregular in the optical may have regular grand-design spiral
arms in the near-IR (Block et al. 1994), which emphasizes the importance of
a new, more dynamical picture of the morphological structure in galaxies.
A step toward that direction is the new dust penetrated classification
of spiral arms in the near-IR (Block $\&$ Puerari 1999; Buta $\&$ Block 2001;
Block et al. 2001), in which bar strength plays an important role.

As discussed by Buta $\&$ Block (2001; BB from hereon) there are many 
quantitative 
parameters which can be used to estimate bar strengths, such as bar-interarm 
contrast (Elmegreen $\&$ Elmegreen 1985) or light remaining after the disk 
and bulge components are subtracted (Seigar $\&$ James 1998). The most 
commonly used method is the maximum ellipticity of a bar, 
an approach justified by the analytical models by Athanassoula (1992),
who showed that the non-axisymmetric forces in the bar correlate with
the bar ellipticity. This method has been recently refined by 
Abraham $\&$ Merrifield (2000), who consider both the inner and outer contours 
of the image to better resolve the ellipticity of a bar. However, the
ellipticity is not a full description of bar strength. In fact, a more
physical approach has been taken by BB
who estimate bar torques by calculating tangential forces in the bar region,
taking into account also the underlying axisymmetric potential. Indeed, when
refined, the bar torque method is probably the most promising way of 
estimating bar strengths. 

When the bar torque method is finely tuned, future refinements will include
the complex bar structures seen in many galaxies; taking more properly
into account the vertical scale heights and their gradients, as well as 
taking into account bulge stretch scenarios upon deprojection of the images.
In the bar torque method 
there are also different ways of evaluating the gravitational potential
and it is important to verify that the different methods
give mutually consistent results. For example, BB used the 2D Cartesian 
integration method by Quillen et al. (1994), whose new contribution in the 
potential evaluation was that the vertical
density profile of the disk was taken into account in the convolution 
function. The potential was calculated in Cartesian grid by applying
Fast Fourier Transform techniques (see also Elmegreen et al. 1989). On the
other hand, in our study of IC 4214 (Salo et al. 1999) we evaluated the 
barred potential by
first ``smoothing'' the image by
calculating the Fourier decomposition of the surface density in a polar grid. 
In principle these two methods should give similar results.

In the present study bar strengths are calculated in JHK-bands for 107 
spiral galaxies using the polar method (Salo et al. 1999) for the evaluation
of the gravitational potential.
The method has been improved  by taking into account the recent 
observational work showing that bars in early-type galaxies are thicker than 
in late-type galaxies. Also, the effects of a distance dependent scale height, 
detected in many boxy/peanut shaped disks, are estimated. The 
algorithm of calculating forces is described, the different ways of 
estimating the gravitational potential are compared and Fourier analysis is
applied for the analysis of bars. Also, biases of the sample are studied
and the distributions of bar torques among the de Vaucouleurs's optical bar
classes are compared. In future the method will be further developed 
to better take into account the observational properties of bars and
bulges in galaxies. The measurements of this paper have been used
in comparisons of bar strengths between active and non-active galaxies
by Laurikainen et al. (2002).

 \section{The sample selection}

A sample of spiral galaxies was selected from the Third Reference Catalogue
of Bright Galaxies (de Vaucouleurs et al. 1991, RC3) requiring that 
$B_T < $ 12.5 mag, cz $<$ 2500 $km \ s^{-1}$ and the inclination 
INC $< 67^0$. As a database we use the 2 Micron
All Sky Survey (2MASS). As only about 50 $\%$ of the
galaxies in our original sample have images available in 2MASS,  
the sample in the present form is not magnitude-limited. 
Some of the weakest images were also eliminated, because bar strengths could
not be measured in a reliable manner for them. The final sample consists of 
107 spiral galaxies, of which 31 are barred (SB) in RC3, 53 show nuclear 
activity, 42 are early-type spirals (SO/a-Sb) and the rest of the galaxies
belong to late-types. To active galaxies
we include Seyferts, LINERs and HII/Starburst galaxies, the type of 
activity being
taken from NASA/IPAC Extra-galactic Database (NED), where the most recent
classifications are available. 

The frequency of bars in our sample is typical in 
comparison
to other samples of galaxies. Including to barred galaxies both SB and
SAB types we find a bar frequency of 62 $\%$, which is similar to that
obtained by Sellwood $\&$ Wilkinson (1993) for field galaxies (60 $\%$ barred)
and by Ho et al. (1997) for a magnitude-limited sample ($B_T<$12.5)
of spiral galaxies (59 $\%$ barred). A somewhat larger fraction of barred 
galaxies has been found by Hunt $\&$ Malkan (1999) for a sample selected by
12 $\mu m$ radiation (69 $\%$ barred) and by Moles et al. (1995) for a 
magnitude-limited sample extending to $B_T$=13 mag (68 $\%$ barred). 
The numbers of SB and SAB galaxies
in our sample are rather similar (29 $\%$ and 33 $\%$). While identifying 
bars in the near-IR by Fourier techniques (see Section 5.3) the fraction 
of barred galaxies was found to be 40 $\%$. 

The fraction of SB galaxies among Seyferts 
and LINERs in our sample is 30 $\%$, being similar as for the whole sample,
in agreement with the bar fractions detected in the samples
by Ho et al. (1997) and Mulchaey $\&$ Regan (1997). 
An enhanced frequency of bars is generally associated to HII/Starburst 
galaxies,
for example, the fraction of SB-galaxies among the Markarian Starburst
galaxies is even 75 $\%$ (Hunt $\&$ Malkan 1999). The bar fraction
among HII/Starburst galaxies somewhat decreases when weaker nuclear 
star formation activity
is considered: namely, in the whole 12 $\mu m$ sample by Hunt $\&$ Malkan the 
fraction of SB-galaxies is 53 $\%$, but while concentrating to 
smaller distances (cz $<$ 5000
$km \ s^{-1}$), it decreases to 46 $\%$. In our sample only 2 out of 17 
HII-galaxies are classified as SB, which can be partly 
understood by the small distances of the galaxies 
($<cz>$= 1200 $km \ s^{-1}$): most probably 
we are picking the lower end of the nuclear HII-luminosity function, 
where bars may not be the driving forces for nuclear star formation.

The frequency of active nuclei in galaxy samples depends strongly on the
redshift range studied: at small distances almost all galaxies show
nuclear activity in some level (Ho et al. 1997), while the number of 
strong 
active nuclei increases with redshift. In our sample 53 $\%$
of the galaxies have active nuclei in terms of Seyferts, LINERs and 
HII/Starburst galaxies. As we have the same apparent magnitude limit
as Ho et al., more HII/Starburst galaxies would be expected. The reason
why we don't have more HII/Starburst galaxies most probably is that NED 
may not sample well the lowest level activity. 
The morphological types of Seyferts and LINERs in our sample
are peaked to Sab galaxies, which is between the mean 
morphological types for Seyfert 1 (Sa) and Seyfert 2 (Sb) galaxies
(Malkan et al. 1998), whereas 3/4 of the non-active galaxies
belong to late Hubble types.

\section{Evaluation of the gravitational potential}

Like in BB, also here the bar strengths are 
estimated from the magnitude of the
non-axisymmetric gravitational perturbation in comparison to the mean
axisymmetric radial force field. For each radius $r$ and azimuth $\phi$ we
calculate the tangential force $F_T= {1 \over r}
\partial\Phi/\partial\phi$ and the radial force $F_R= \partial \Phi/
\partial r$, and define the relative strength of the perturbation as
          $$Q_T (r) = {F_T}^{max}(r) / <F_R(r)>,\eqno (1) $$  

\noindent 
where the average of radial force over azimuth is taken. Following BB the
tangential force maxima were calculated separately in four quadrants
of the image and the mean of these in each distance is used as the
maximum $F_T$. In order to obtain a single measure for the strength we
use $Q_b$, which is the maximum $Q_T$ in the bar region. The radial
distance where this maximal perturbation takes place is denoted by
$r_{Qb}$.

For the force calculation the gravitational potential $\Phi$ in the bar
region must be evaluated. We assume that the surface density $\Sigma$
is proportional to the surface brightness obtained from the near-IR
image, and that the vertical density distribution follows a model
profile $\varrho_z(z)$, normalized to unity when integrated over
$z$. First it was assumed that the z-dependence of the mass density
obeys the same formula everywhere in the galaxy (see Section 4.1), in which 
case the
gravitational potential in the central plane of the galaxy can be
written as (e.g. Quillen et al. 1994)
          $$\Phi (x,y,z=0) = -G \int_{-\infty}^{\infty} \int_{-\infty}^{\infty} \Sigma (x', y') g(x-x', y-y')dx'dy' \eqno (2)$$ 

\noindent where the integral over z-direction is included to the 
convolution function $g(x-x',y-y') \equiv g(\Delta r)$,
with $\Delta r^2=(x-x')^2+(y-y')^2$, defined as 
       $$g(\Delta r) = \int_{-\infty}^{\infty} \varrho_z(z) (\Delta r^2+z^2)^{-1/2} dz .\eqno (3)$$

Several models for $\varrho_z(z)$ were applied, including the
exponential model with scale-height $h_z$,
          $$\varrho_z(z) = {1 \over {2 h_z}} exp (-|z/h_z|),\eqno (4)$$ 

\noindent and the often used isothermal sheet model (van der Kruit $\&$ 
Searle 1981),
          $$\varrho_z(z) ={1 \over {2 h_{sech2}}}  sech^2(z/h_{sech2}), \eqno (5)$$ 

\noindent where $h_{sech2}$ is the isothermal scale-height.
Usually, when isothermal model is used, $h_{sech2}$ is set equal to
$2h_z$, to yield a same slope at large $z$ as for the exponential
model. According to van der Kruit (1988) galactic disks may, however,
deviate from isothermal shape near the galactic plane, in which case a
better approximation for the density distribution would be
          $$\varrho_z(z) = {1 \over {\pi h_{sech}}} sech (z/h_{sech}).\eqno (6)$$ 

\noindent In the case of $sech$-law, Barnaby $\&$ Thronson (1992) 
identify $h_{sech}$ with $(\pi/4) h_z$,
to get the same central plane density as in the $sech^2$-law.
However, in this study we briefly check $sech$-law with $h_{sech}=h_z$, in which case
the slope of the vertical profile corresponds to exponential model at large $z$.
\noindent De Grijs et al. (1997) have explored a more general family of 
fitting functions, which include the above functions as special cases. 
As will be discussed below the adopted vertical model
has rather little influence on the derived forces, provided that
appropriately defined scale-heights are used. 

In comparison to BB we use a somewhat different method for the
evaluation of the gravitational potential, mainly because the 2MASS
images used here have a more poor resolution (1''/pix) and lower
signal-to-noise (S/N) ratio than the images used by BB. Thus, instead 
of calculating the gravitational
potential directly from the image pixels, the images are first
``smoothed'' by calculating the Fourier decompositions of the surface
densities in different radial zones,
$$ \Sigma (r,\phi) = \Sigma_0(r)+ \sum_{m=1}^{m=\infty} \Sigma_m(r,\phi) =  
A_0 (r) + \sum_{m=1}^{m=\infty} A_m(r) \cos \left[m \left(\phi -\phi_m(r) \right) \right]. \eqno (7)$$ 

\noindent The density of each
Fourier component is then separately tabulated as a function of radius
and azimuth, and for each density component the corresponding
potential component is obtained by
       $$\Phi_m (r,\phi,z=0) = -G \int_{0}^{\infty} r' dr' \int_{0}^{2\pi} \Sigma_m(r', \phi') g(\Delta r) d\phi',\eqno (8)$$

\noindent with ${\Delta r}^2= {r'}^2+r^2-2rr'cos(\phi'-\phi)$. The integration
over azimuthal direction is carried out by Fast Fourier Transform (FFT),
whereas in the radial direction a direct summation is used. An
azimuthal offset of one half bin is used between density and potential
locations, and the force components at density locations are obtained
from the potential by numerical differentiation. For test purposes, we
also applied Cartesian potential evaluation, solving Eq. (1) with 2D FFT
as in BB. In comparison to direct Cartesian force evaluation
the advantage of our polar method is that also rather weak and noisy
structures can be measured in a reliable manner, due to the smoothing
implied by Fourier decomposition. Also, it is possible to limit the
density to even components, most likely to characterize the
non-axisymmetry related to the bar. Our method also gives directly
the different Fourier modes of the potential and force components,
which are sometimes of interest. 

Like in BB and Quillen et al. (1994), we made use of the fact that the
convolution function $g(\Delta r)$ can be numerically integrated and
tabulated as a function of $\Delta r/h$, $h$ denoting the vertical
scale-factor used, so that the integration over $z$-direction can be
replaced by a much faster interpolation from pre-calculated tables.
In addition, since in the polar method only the integration over
azimuth is carried out by FFT, a distance-dependent $h$ can be used: in
this case $g(\Delta r / h)$ in Eq. (8) remains cyclic with $\phi' -
\phi$ even if $h=h(r')$. In the Cartesian 2D FFT this can not be done
as the convolution function needs to be cyclic in both $x-x'$ and
$y-y'$.

In Fig. 1 we display
the convolution functions corresponding to various vertical models.
The upper row shows the density profiles and convolution
functions corresponding to exponential, $sech$, and  $sech^2 $-model profiles.
The difference in $g(\Delta r)$ is significant only for $\Delta r < h_z$. In
the lower row of the figure we illustrate the relative insensitiveness of 
$g(\Delta r)$ for a wider range of $\varrho_z(z)$ models, including the
Gaussian model,
          $$\varrho_z(z) = {1 \over {\sqrt{2\pi} h_{gauss}}} \exp \left(-{z^2 \over 2 {h_{gauss}}^2}\right),\eqno (9)$$

\noindent and a uniform slab-model,
          $$\varrho_z(z) = {1 \over {2 h_{uni}}}, \ \ -h_{uni} < 
z < h_{uni}.\eqno (10) $$

\noindent In order to make the models 
comparable, the scale factors
in each case were chosen in a manner that yields the same vertical dispersion as the exponential model with
$h=h_z$, namely
$<z^2\varrho_z(z)>/<\varrho_z(z)>=2{h_z}^2$. In this case
$$h_{gauss}/h_z = \sqrt{2},$$
$$h_{sech2}/h_z = \sqrt{24}/\pi,$$
$$h_{sech}/h_z = \sqrt{8}/\pi,\eqno (11)$$
$$h_{uni}/h_z = \sqrt{6}.$$

\noindent As expected, the more peaked the density profile is toward the central plane, the 
larger is the value of $g(\Delta r)$ when $z \rightarrow 0$. In each case $g(\Delta r) \propto
1/\Delta r$ for large $\Delta r$,
while $g(\Delta r)$ increases logarithmically when $\Delta r \rightarrow 0$.
In the case of a uniform slab, analytical integration yields
$$g_{uni} = {1 \over h_{uni}} \log \left({h_{uni}+\sqrt{h_{uni}^2 + {\Delta r}^2} 
\over {\Delta
r}}\right),\eqno (12) $$

\noindent in agreement with the result of the numerical integration shown in 
Fig. 1. For comparison, Fig. 1 displays also the case of using 2-D 
softened gravity,
$$g_{soft}=1/\sqrt{{\Delta r}^2+\epsilon^2}, \ \ \rm{with} \ \epsilon^2=2h_z.\eqno (13)$$

In conclusion, the convolution function depends very little on the model 
used for the
vertical mass distribution as long as models with a same vertical dispersion 
are compared. In the next section bar strengths
calculated using different vertical models
are compared, as well as results obtained by the two different integration 
methods.

\section{Testing the algorithm}

\subsection{Polar vs. Cartesian integration}

We next investigate our algorithms of force evaluation using NGC
1433 as a test case. For this galaxy Ron Buta has provided us with his
high quality H-band image, with a 1.141 arcsec pixel scale (Buta et al.
2001). Fig. 2 compares our
standard polar method, and our Cartesian potential evaluation, where
the de-projected image is interpolated to a density array and then
FFT in Cartesian coordinates is applied with a grid
resolution ranging from $64 \times 64$ to $512 \times 512$. In both
polar and Cartesian cases the forces where calculated from a region
with radius of $r_{max}=180$ pixels (205 arcsecs). According to
Fig. 2 the obtained $Q_T$-profiles converge rapidly as more
Fourier components are included, or when a finer Cartesian grid is
used. The profiles obtained with the two methods also agree well: the
maximum $Q_T$ agrees within a few percent. The only differences
appear at large $r$, where the polar method yields smaller $Q_T$. This is
due to distance-dependent smoothing implied by the polar method, whereas
in the
Cartesian integration the maximum tangential forces for large $r$ are
mainly due to spurious values connected to individual image pixels. Thus,
since the outer parts of the galaxy have a low S/N-ratio, the smoothing
in the polar method is also physically well motivated. 

As the polar method is based on calculating Fourier amplitudes for
different density components, it is interesting to check how much each
component contributes to the total force: in Fig. 2 even components
up to m=10 are studied. We have used an exponential vertical profile,
both with $h_r/h_z=2.5$, representing an early-type galaxy,
and with $h_r/h_z=10$, to study the effect of assuming a very flat
disk. The density amplitudes m=2 and m=4 are known to be the
strongest in bars, while in early-type galaxies the m=6 and m=8 may also be
significant (Ohta 1996). For NGC 1433 (type Sb) the density amplitudes
we find in H-band are fairly similar to those displayed for the I-band by
Buta (1986), with $A_2/A_0$ attaining a maximum $\approx 0.97$ at
$r=110''$, and a secondary maximum of $\approx 0.6$ at $r\approx 50''$. For
$m=4, 6, 8, 10$ the maximum $A_m/A_0$ are about $0.42, 0.22, 0.13,
0.09$, respectively, and for the resulting force the maximum appears at
$r_{Qb}=70''$, which corresponds to about $90\%$ of the bar radius
($R_{bar}=77''$, Buta 1986). For $h_r/h_z$=2.5 about 75$\%$ of the tangential 
force is due to the m=2
density component, and with the inclusion of the m=4 component $Q_b$
increases to 97 $\%$ of its value obtained by including all even
components up to m=10. However, for the case of a flatter galaxy the
influence of higher order Fourier components would become somewhat
more important: in the case of $h_r/h_z=10$, $Q_b = 70\%,
90\%, 98\%$ of its maximum value, when including the density components
up to $m_{max}=2, 4, 6$, respectively. Thus even in this case the density
components above m=6 have fairly little influence. This difference in
the relative importance of various Fourier components follows from the
fact that planar density variations correspond to force variations
only if their planar scale significantly exceeds $h_z$: smaller
$h_r/h_z$ thus suppresses the force variations corresponding to large $m$.

In the above example the calculation region covered well
the optical disk of the galaxy, and the maximum $Q_T$ was
obtained at $r_{max}/r_{Qb} \approx 3$. However, in the case of 2MASS
images the outer disks for distant galaxies are sometimes not deep enough. 
In order
to check the influence of the size of calculation region, a series of
integrations with decreasing $r_{max}$ was conducted for NGC 1433. For
$r_{max}/r_{Qb} =2$ the resulting value of $Q_b$ was only about $2\%$
smaller (no shift in $r_{Qb}$), whereas for $r_{max}/r_{Qb} =1.5$ the 
reduction was already
significant, amounting to $15\%$. About $2/3$ of this reduction is due to
reduced tangential force, while the remaining part comes from the
overestimated radial force due to disk truncation. However, the location
of $Q_T$-maximum was still fairly little affected (66'' vs. 70'').
All our subsequent force evaluations from 2MASS images are based on
images covering a radius at least twice as large as the derived $r_{Qb}$.

We also checked the influence of different model functions
for the vertical profile on NGC 1433 (Fig. 3). The models correspond
to those studied in Fig. 1, each having the same vertical dispersion.
As expected, based on the behavior of the convolution functions, a more
centrally peaked vertical profile yields a slightly larger $Q_b$, although 
the shape of the
$Q_T$-profile is little affected. For  
early-type galaxies with $h_r/h_z=2.5$, the difference in $Q_b$ between
the exponential and uniform models  
is $15\%$, which is much less than the difference between the cases
$h_r/h_z=2.5$ and $h_r/h_z=10$ for a fixed vertical model function.
Also, while 
comparing more realistic vertical models, namely the exponential and 
the isothermal model, $Q_b$ is affected only $5\%$.
In general, when the vertical extent of the disk is reduced, the exact 
form of the vertical density law becomes less important. 
In the limit of very large $h_r/h_z$ the 
difference
between including 3D vertical profile and the use of softening
in 2D force-evaluation becomes also small, in agreement with Salo et al. 
(1999). However, for large disk thicknesses ($h_r/h_z$ = 2.5), the use of 
softened
gravity severely underestimates $Q_b$. 

\subsection{Distant dependent disk thickness}

As many galaxies have boxy/peanut shaped structures, for which the vertical 
thickness has found to increase towards the outer parts of the galaxies,
we also studied a case with distant dependent
vertical scale-height. Radial 
distributions of $Q_T$ were calculated for NGC 1433 using an exponential
density model, when $h_z$ was assumed to have a mean value and a gradient
typical for early-type galaxies (see Section 5.1). As shown in Fig. 4 
the vertical density gradient has a fairly small effect on bar strength: 
in comparison to the case where a constant $h_z$ was assumed, a gradient
of 0.05 in $h_z$ affects $Q_T$ at maximum only by $4\%$. In the figure three
cases are demonstrated showing how the location of the mean vertical thickness
affects $Q_T$ measurement. In comparison to constant vertical scale-height,
radially increasing $h_z$ typically slightly reduces the derived bar strength. 
This is due to two
competitive effects: on one hand the tangential force increases in the 
region where the vertical scale-height is reduced, but at the same time 
axisymmetric radial force is also increased in the same region (see 
Fig. 5).

In interacting galaxies gradients of the vertical scale-heights are 
generally larger than for normal galaxies, and contrary to normal galaxies,
may also appear in late-type systems. Therefore we studied also some cases,
where the $h_z$ gradient was twice as large as for normal galaxies. For
$h_r/h_z$=2.5 the effect of the gradient on $Q_b$ 
was still unimportant $Q_b$ being reduced only by $3\%$. Thin disks were 
even more less affected: for $h_r/h_z$=10
we found that while adding a gradient of 0.03, $Q_b$ was maintained unchanged. 
We can safely conclude that the vertical scale-height gradient, 
typical for boxy/peanut shaped bars, does not significantly affect bar 
strength.

\subsection{Comparison to BB}

We next compare our results with those obtained by BB. Bar strengths for
13 galaxies in the sample by BB were calculated in H-band, using
the same orientation parameters as used in BB (see Table 4). Also,
following BB the vertical mass distribution was approximated by
an exponential function with $h_z$ = 325 kpc, which is the scale-height
of the disk in the Milky Way. As shown in Fig. 6, there is a small
shift toward somewhat larger $Q_b$ in our measurements, but generally the 
agreement between the two measurements is quite good. No softening was 
used in our potential evaluation. A possible cause of the small 
difference is that some form of additional gravity softening was 
included to the potential evaluation in BB, besides using a vertically
extended density model. This possibility is suggested by the form of the
convolution function $g(r)$ shown in Fig. 2 in Quillen et al. (1994),
being the method used by BB. In comparison to our $g(r)$ (Fig. 1), their
function seems to attain a constant value for $\Delta r / h$ -$>$ 0,
much like our curve for softened gravity. Nevertheless, no explicit
gravity softening was mentioned in Quillen et al., so it is not clear
whether this is the cause for the difference.
It is also worth
noticing that the comparison was made using 2MASS images, many of which
were of more poor quality than the images
used by BB. In Table 4 we also show that the vertical 
density model barely affects the derived bar strength.

\section{Calculation of the non-axisymmetric forces for the sample}

\subsection{Observed parameters in the potential evaluation}

As discussed in the previous section,
several approximations were made while calculating the gravitational 
potential, largely following BB.
The main assumption was that the mass-to-luminosity ratio (M/L) is  
constant throughout the disk.
This assumption was made for simplicity, because otherwise possible M/L-
variations along the disk should be known for each individual galaxy.
The studies of color gradients in galaxies have shown
that the central regions of the disks
are often redder than the outer disks, which actually suggests larger stellar
M/L-ratios in the inner disks (Bell $\&$ de Jong 2001). 
Another indicator of a possible non-constancy of the
M/L-ratio comes from the comparison of the optical surface photometry with
the surface mass densities obtained from the rotation curves, showing that
the M/L-ratio may vary along the disk (Takamiya $\&$ Sofue 2000). 
However, it is possible that the M/L-ratio is rather   
constant in the bar region (see Quillen et al. 1994), in which case the 
made assumption should be reasonably good. 

In the evaluation of the barred potential a model for the vertical 
mass distribution was assumed.
The most commonly used models are the exponential and isothermal 
functions, both being
physically justified. An isothermal
density function is expected if stars, once formed, do not interact with 
the other components of the galaxy. Stellar populations of different 
ages can then be understood as quasi-independent components with
different velocity dispersions (Dove $\&$ Thronson, 1993) 
so that a more complete picture 
would be achieved by assuming a super-position of a large number of
isothermal sheets with different $h_z$ (Kujken 1991). On the other hand, if 
gas settles 
into an equilibrium prior to star formation or considerable heating
of the disk has occurred during the life of a galaxy, an 
exponential density profile is expected (Burkhead $\&$ Joshii 1996). 
However, empirically the two functions 
are difficult to distinguish, because the density functions look
similar at large vertical heights, whereas near to the galactic plane
the evaluation of the density function is complicated by the effects of dust.  
For the vertical density distribution we used an exponential function.
However, based on discussion in Section 4.1, the uncertainty in the 
vertical model function should have fairly small influence on derived forces.  

For bar strength measurements, more critical is the thickness of the disk. 
It has been found to depend on the 
morphological type of the galaxy (de Grijs 1998) and also to correlate 
with the 
radial scale-length of the disk (Wainscoat et al. 1989; van der Kruit 1988).
For early-type galaxies we used  $h_r/h_z$=2.5, 
and for late-type systems $h_r/h_z$=4.5. For $h_r$ we use the optical 
V-band scale-lengths from Baggett 
et al. (1998), and if not available, they were estimated from the 2MASS 
images by us. Optical scale-lengths were used, but they are expected to be 
rather similar with those in the near-IR: namely using the bulge-to-disk 
decompositions given by de Jong (1996) for 186 spiral galaxies, we found 
that the scale-lengths in V-band deviate on the average only 5$\%$ from 
those in K-band. For three of the galaxies, pgc 10266,
pgc 15821 and pgc 40097, the scale-lengths by Baggett et al. were judged
unrealistic: for two of the galaxies they were rather measures of the 
brightness slopes in the bulge region, and for one galaxy the given 
scale-length represented 
the outermost very shallow part of the disk, while we are interested in
the disk under the bar. Therefore, also for these three galaxies the 
scale-lengths were estimated from 2MASS images. 

Bars and bulges have often rather complex structures and may in some 
cases be difficult to distinguish from each other. For example,
45 $\%$ of all galaxies (SO-Sd) may have boxy or peanut shaped bulges
or bars, having vertical disk profiles that become thicker 
towards the outer parts of the disks (Lutticke et al. 2000, Schwarzkopf 
$\&$ Dettmar 2001).
These gradients are pronounced in early-type galaxies, but generally 
do not appear in late-type galaxies.
The boxy/peanut structures are often thought to be bulges, but resent 
observations rather support the idea that they more likely represent
the thick parts of the bars. This interpretation is also supported by
the simulations by Athanassoula (2002) who has shown that boxy/peanut 
structures are formed from the particles of the disks or bars during
the evolution of the galaxy. In this work we have made the simplified
assumption that $h_z$ is constant throughout the disk, although the 
effect of the thickening of the disk toward the outer parts of the disk
is also investigated. In the test we used a $dh_z/dr$ gradient of 0.05, typical
for normal SO/a galaxies (Schwarzkopf $\&$ Dettmar 2001). Galaxy interactions
and mergers of small satellite galaxies can also be efficient in
thickening the disk especially in the outer parts of the galaxies (Toth $\&$
Ostriker 1992; Walker et al. 1996). In fact, observations by Schwarzkopf
$\&$ Dettmar have shown that the vertical heights of the disks for 
interacting galaxies can be twice as large as for non-interacting
galaxies. Bulges were assumed
to be as flat as the disks, which is also a simplified assumption, but
may still be valid for the triaxial bulges of SB-galaxies (Kormendy 1982, 
1993). Evidently, the treatment of bulges need to be improved in future.

\subsection{Calculation of forces for sample galaxies}

As a database JHK images of the 2MASS survey were used.
The spatial resolution of the images was one arc second, and
the image quality was generally best in the H-band. 
Our procedure of estimating
the non-axisymmetric forces consisted of the following steps: (1) 
cleaned sky-subtracted mosaics were constructed, (2) galaxies were de-projected
to face-on orientation, (3) the images were
rebinned by a factor of two, (4) Fourier decomposition of the surface 
density was calculated, and barred potentials were evaluated using the 
even components
up to m=6, (5) the tangential ($F_T$) and the mean axisymmetric radial
forces ($<F_R>$) were calculated, and finally (6) maps of the force ratios 
were constructed: 
 
	$Q_T(r, \phi) = F_T(r, \phi)$ / $ <F_R(r, \phi)> $.

\noindent For a bar the map shows four well defined regions where 
the force ratio reaches a maximum or minimum around or near the end of the bar.
As in BB, we call this structure as a ``butterfly pattern''.

The field of view in the 2MASS images is relatively small 
and also a large fraction of 
the galaxies in our sample are quite nearby objects so that
mosaics of 2-5 images were generally made.
The image quality was not as good in the borders as in the
central parts of the frames, which 
in principle could seriously affect the quality of the mosaics in the
regions of interest. For the most nearby galaxies this was
not a problem, because the S/N-ratio was high in all parts 
of the images. For the distant objects the problem was solved so 
that we never 
combined images if their borders appeared in the bar region 
or in the central parts of the galaxies. Before combining the images
the over-scan regions were removed and the sky values given in the image 
headers were subtracted. The background
levels of the frames were then refined to get similar
count levels in the galaxy regions in the combined images.
Positioning of the 
frames in the mosaics was done using stars common in the combined images.
Finally the foreground stars and bad pixels were rejected.

The cleaned (mostly mosaics) images were de-projected to face-on 
orientation using the position angles (PA) and inclinations of the disks
given in Table 1. In the table some other properties of the galaxies 
like the mean revised morphological type $T_m$, the blue apparent 
magnitude $B_T$, the type 
of nuclear activity, and the scale-length of the disk $h_r$, are also 
shown. The orientation parameters, the morphological types and the apparent
magnitudes are from RC3 if not otherwise mentioned.
For some of the galaxies, 
instead of using the orientation parameters from RC3, they were 
estimated from the Digitized Sky Survey Plates by us.
For example, for pgc 18258 the surface brightness contours are  
clearly affected by the super-position of a small companion, which has 
not been taken into account in the orientation parameters in RC3.
For the galaxies that had no estimation of PA in the literature,
but the disks were in nearly face-on orientation, we used INC=0.
In order to estimate properly
the axisymmetric radial forces the bulges were not subtracted. 

The maximum of non-axisymmetric forces, $Q_b$, and their radial distances, 
$r_{Qb}$,
are shown in Tables 2 and 3, measured in J, H and K-bands. The uncertainties
attached in the tables are the maximum deviations between the four image 
quadrants. However, the largest uncertainty (see Section 4.1) is  
due to the observed
scatter in $h_r/h_z$ within each morphological type, which for example for
Sc-galaxies induces an uncertainty of about 
15 $\%$ in $Q_b$.
Non-axisymmetric forces are also sensitive to the orientation parameters 
of the galaxies: BB estimated that an uncertainty of $\pm$ \ 10 $\%$ in the
inclination and position angle can induce an uncertainty of two bar strength
classes (one bar strength class corresponds to 0.1 units in $Q_b$).
Indeed, in future this work can be improved when accurate
position angles and inclinations will be available for all of the galaxies 
studied. Especially for interacting galaxies the photometrical 
orientation parameters are generally determined
from warped or distorted outer disks so that 
the kinematic observations for them give more reliable values. 
For example, for M51 the kinematically (Tully 1974) 
and photometrically (Spillar et a. 1992) estimated inclinations 
deviate by about $20^0$, and for IC 4214 by $10^0$ (Buta et al. 1999), for
which the difference can be explained by the bar potential (see Salo 
et al. 1999). On the other hand, the uncertainty caused by the vertical model 
($exp$ v.s. $sech^2$) is negligible for thin disks  
and only about 5 $\%$ for thick disks. Also,
the number of Fourier modes or the size of the measurement region do not 
affect $Q_b$ significantly, if large enough measurement regions 
($r_{max}$ $>$ 2$r_{Qb}$) and enough Fourier modes (even modes up to m=6) 
are used. The effect of the boxy/peanut shaped structures
in terms of increasing vertical scale-height toward the outer parts of
the disks also appeared to be insignificant for bar strength.

\subsection{Identification of bars and calculation of bar ellipticity} 

Bars in the near-IR were identified by Fourier techniques. In distinction 
to bars in general, we call ``classical bars'' the morphological 
structures with the ratio of the Fourier amplitude $A_2/A_0$ larger than 0.3, 
and with the m=2 phase maintained nearly constant in the bar region (Table 2). 
For these 
bars the m=4 amplitudes are also pronounced, there is a clear maximum in 
the $Q_T$-profile in the bar region and the ``butterfly pattern'' shows 
four regular structures. The $Q_T$-profiles and the ``butterfly patterns'' 
for these galaxies are shown in Fig. 7, showing also the
m=0,2, and 4 surface brightness profiles. The length of the 
region with a constant m=2 phase was taken to be a measure of the bar 
length. Both the m=2 and m=4 phases are maintained nearly constant in the bar 
region, but in some cases the m=4 amplitude drops at a slightly shorter 
distance, in which case we used the mean length of the constant m=2 and m=4
phases as the bar length. 

In addition to the ``classical bars'', other bar-like structures and 
non-axisymmetries are identified in many of the galaxies in our 
sample. 
Actually a large majority of galaxies have 
non-axisymmetric forces, manifested as maxima in the $Q_T$-profiles, whereas
in 20 $\%$ no non-axisymmetric forces were detected 
above the background level ($Q_b$=0.01-0.1). Due to differences in image 
quality no single 
minimum $Q_b$-value defines the bar-like potential.
The non-axisymmetric forces presented in Table 3 deviate from the ``classical
bars'' in a sense that the m=2 phase is not maintained constant in the
assumed bar region and the regular ``butterfly patterns'' do not necessarily 
appear in the force field.
In 15 $\%$ of the galaxies in our sample the $Q_T$ maxima are manifestations 
of strong spiral arms in the outer parts of the galactic disks.

As discussed by BB, both spherical and flattened bulges can  
affect 
$Q_b$-measurements: in a case of an intrinsically spherical bulge, the effect
of assuming a bulge as thin as the disk is to overestimate the axisymmetric 
radial force and 
consequently to underestimate the relative bar strength. On the other hand,
while de-projecting the image to face-on orientation bulges might cause
artifacts in the direction of the minor-axis of the disk.
The problem of large bulges was here avoided by limiting
to those cases where the maximum tangential force appeared
outside the bulge region.
The small bulges still makes it difficult 
to detect mini-bars, but by
subtracting the m=0 component and by taking into account the de-projection 
effects,
mini-bars could be detected for pgc 10488, 33371, 37999,
40153 and 43495, being previously identified also by 
Buta $\&$ Crocker (1993), Perez-Ramirez et al. (2000),  Knapen et al. (1995), 
and by Block et al. (1994).

Based on the analytical work by Athanassoula (1992) the maximum 
ellipticity of a bar can be used as an approximation of bar strength, and 
its radial distance as an estimate of bar length. We calculated the 
ellipticity profiles using a method described in Laurikainen $\&$ Salo (2000)
in which ellipses were iteratively fitted to 
the isophotes of the surface brightnesses. 
The maximum ellipticities $\epsilon$ and their radial distances 
$r_{\epsilon}$ in the bar region for the 
``classical bars''are shown in Table 2 .
In Laurikainen et al. (2002) these ellipticity measurements were utilized
to show a good correlation between $Q_b$ and $\epsilon$.

\section{The sample biases}

The measurements reported in this work
are used to compare bar strengths of the active and
non-active galaxies (Laurikainen et al. 2002), so that it is important
to study possible biases between different subgroups in the sample.

Active galaxies in our sample appeared to be somewhat brighter than the
non-active systems, which is illustrated in Fig. 8a, showing also a weak 
correlation between the bar length and the absolute blue magnitude $M_B$, 
of the galaxy. However, while scaling the bar length to the
scale-length of the disk, the bias was largely diluted (Fig. 8b).
A similar correlation has been found previously by Kormendy (1979)
for optically measured bar lengths.
Due to the magnitude bias longer bars may have been selected for the active 
galaxies in our sample.  
However, we confirmed that this bias does not affect the mean bar strengths
in the compared subsamples. This was checked by dividing the non-active
galaxies to two magnitude bins with $M_B$ larger and smaller than
-19.8 mag, resulting to practically identical mean forces, 
$<Q_b>$ = 0.22 $\pm$ 0.10 and 0.24 $\pm$ 0.12,
respectively (the uncertainties indicate the sample standard deviations).

The relative number of barred galaxies rapidly drops in our sample
when the 
inclination of the disk is larger than $50^0$ (see Fig. 9). This is the
case both for SAB and SB-galaxies and for the 
``classical bars'' identified in the near-IR. 
The inclination distribution for the galaxies in the 
whole sample largely follows a random distribution
of orientations, dN/di $\sim$ sin i, expected for an unbiased sample.  
Again, since in Laurikainen et al. we compare bar strengths between different 
subsamples it is important 
that this inclination bias does not affect the mean $Q_b$-values.
It appeared that $Q_b$ does not correlate with 
the inclination of the disk (see Fig. 10): especially for the
galaxies with no ``classical bars'' $Q_b$ is similar
for all inclinations. For the galaxies with ``classical bars''
there may be a lack of strong bars among the nearly
face-on galaxies (INC $< 25^0$) and clearly among the highly inclined galaxies 
(INC $> 60^0$),
which is probably a manifestation of the uncertainties in the inclination
determinations. 

In order to check whether the dependence of bar detection frequency on 
inclination is specific for our sample or does appear also  
in larger samples of galaxies, we picked up all galaxies
brighter than 15.5 mag from RC3 and constructed histograms 
for SA, SAB and SB-galaxies (Fig. 11), similar to those in Fig. 9. 
Evidently, a similar bias appears also
in RC3: for SA-galaxies the inclinations are well sampled to $60^0$,  
for SB-galaxies to
$50^0$ and for SAB-galaxies the limit of well sampled galaxies is even 
lower than that. 
Quite surprisingly, the
number of SAB-galaxies at inclinations larger than $40^0$ drops much more
rapidly than the number of SB-galaxies indicating that classification 
of a galaxy as SAB is very ambiguous. It also seems
that in RC3 there is a deficiency of galaxies with very low assigned 
inclinations 
(less than $20^0$), supporting the above interpretation that the lack
of strong bars in nearly face-on galaxies is solely due to uncertainties
in the orientation parameters. The found inclination bias might be of 
importance 
for example when bar frequencies are compared between low and high
redshift galaxies. In order to study a possible redshift dependence of 
the bias, 
barred galaxy fractions in two inclination bins were compared at two 
magnitude intervals in RC3.
The bias was found to become more significant toward the fainter galaxies:
the relative number of SB galaxies with INC $> 50^0$ dropped from 32 $\%$
to 25 $\%$ while going from the magnitude interval $B_T < $ 12 mag to 
$B_T$= 13.5 - 14.5 mag, whereas the relative SB galaxy numbers 
with INC $< 50^0$ were identical (38 $\%$) in the two magnitude intervals. 
This means that when 
bar frequencies at high and low reshifts are compared, bar frequencies
of distant galaxies are easily underestimated.

\section{Comparison of optical and near-IR bars}

The distributions of the non-axisymmetric forces among the de Vaucouleurs's
(1963) optical classes SA, SAB and SB are shown in Fig. 12 in J, H and 
K-bands. In agreement
with BB and Block et al. (2001), the overlap between the different
de Vaucouleurs's types is
significant. Based on estimating the ellipticities of bars, a somewhat 
different result was obtained by Abraham
$\&$ Merrifield (2000), who argued that SB-galaxies are clearly separated
from SA and SAB-galaxies in bar strength. In our sample SA and SAB galaxies have
fairly similar non-axisymmetric forces, the only difference being that some
SA galaxies have no non-axisymmetric forces. 

Actually, it seems that SB-galaxies can belong to any of the bar strength
classes from 1 to 6, as defined by BB. In our sample the minimum $Q_b$=0.09
for SB galaxies is largely due to the image quality. We find that 
bars in SB galaxies are very similar to the ``classical bars'' in the near-IR:
namely 95 $\%$ of all SB galaxies in our sample are classified as 
``classical'' in the near-IR. They also have very similar bar 
strength distributions in Fig. 12. However, there are 30  $\%$ 
more ``classical bars'' than SB bars, the excess being distributed to all 
bar strengths, which means that even strong bars can be
hidden by dust in the optical. In addition to bars significant
tangential forces can be induced also by the spiral arms, especially in the
outer parts of the disks, of which M51 
(pgc 47404) is a good example. The spiral related forces 
may in some cases amount even to $Q_b$=0.26, corresponding to a bar 
strength class 3. This
is quite interesting, because it means that even spiral arms 
can induce non-axisymmetric forces in the level typically associated
with moderately strong bars. These non-axisymmetries may have important
implications in secular evolution in galaxies, as for example for the 
onset of near nuclear star formation.

\section{Conclusions}

Non-axisymmetric forces are calculated for 107 spiral galaxies in J,H and K
bands using a  
method where gravitational potentials are evaluated in a polar grid.
Non-softened convolution
function is applied and the vertical distribution of matter is approximated 
by an exponential function. The vertical scale-height of the disk is
taken to be a certain fraction of the radial scale-length of the disk,
and this ratio is assumed to be larger for the early than for the late-type 
galaxies. The M/L-ratio is assumed 
to be constant throughout the disk. The vertical mass 
distribution is generally assumed to obey the same formula everywhere 
in the galaxy,
but tests were also performed to estimate the effect of radially non-constant
vertical thickness.
The phases of the Fourier density amplitudes are used to estimate the 
lengths of the bars.
One of the main concerns of this study is to verify that the different methods
of calculating the gravitational potential give mutually consistent results, 
most of the tests being carried out using a high quality H-image
of NGC 1433 (Buta et al. 2001). In comparison to BB
our method is more suitable for weak and noisy images. Also, it is 
possible to limit only to even Fourier decompositions, most likely 
to characterize the non-axisymmetry related to the bar.
Likewise in polar method it is easy to study distant-dependent $h_z$.
The isophotal ellipticities of bars are also estimated, to facilitate
comparisons to bar strengths estimated from maximal forces. 

The main results are the following:

(1) Cartesian and polar grid methods for the potential evaluation are 
compared. In the Cartesian method the image is sampled to a density
array and then 2D FFT in Cartesian coordinates is applied. In the polar
grid method Fourier decomposition of density is calculated in a polar grid
using FFT in azimuth and direct summation over radius.
We found that similar results are obtained by these two methods for good
quality images, provided that enough Fourier components (up to m=6) are
included, and the resolution of the Cartesian grid is sufficiently large.

(2) Bar strength is found to be rather insensitive to the vertical mass
model of the disk, as long as a same vertical dispersion is assumed for
all models (e.g. $h_{sech2}/h_z = \sqrt{24}/\pi, $ $h_{sech}/h_z = 
\sqrt{8}/\pi$). Boxy/peanut shaped structures, in terms of non-constant 
vertical
scale-heights along the disk, were also found to be quite unimportant for
the evaluation of bar strengths. These parameters affect $Q_b$ less than
5$\%$. The largest uncertainties in $Q_b$ are associated to the large
scatter in the observed vertical scale-height of the disk within one Hubble
type, and to the observed uncertainties in the orientation parameters of
the disks, which both may induce uncertainties of about 10-15 $\%$ in $Q_b$.

(3) Significant non-axisymmetric forces ($Q_b>$0.05) are detected 
in 80 $\%$ of the galaxies in our 
sample. In most cases they were interpreted as bar-like features, based on
significant m=2 Fourier amplitudes in the bar regions and distinct ``butterfly 
patterns'' in the $F_T/<F_R>$-ratio maps. In 40$\%$ of the galaxies 
``classical bars'' were detected, determined as having $A_2/A_0 > 0.3$  
and the m=2 phases maintained nearly constant in the bar region. In some
of the galaxies significant non-axisymmetric forces were detected in the 
outer parts of the disks connected to spiral arms, corresponding even to 
bar strength class 3.

(4) We confirm the previous result by BB and Block et al. (2001) showing the 
large overlap
in bar strength between the optical SA, SAB and SB classes.
Actually, SB-galaxy can belong to any of the bar strength classes between 1
and 6.

(5) We found that 95 $\%$ of SB galaxies in our sample belong to the
``classical bars'' identified in the near-IR, which means that the 
bars are similar. In the optical 1/3 of the ``classical bars'' are 
not classified as SB. Even bars which are in the optical obscured by dust
and which become dust penetrated in the near-IR, cover all bar
classes from 1 to 6, thus indicating that even strong bars can be 
obscured by dust.

(6) Bar lengths are estimated from the phases of the m=2 and m=4 Fourier
components of density requiring that the phase is maintained nearly constant in the 
bar region. Bar length is found to correlate with the galaxy brightness
$M_B$, confirming the previous result by Kormendy (1979) in the optical
region. 

(7) The number of SB-galaxies in our sample drops rapidly at
inclinations $> 50^0$. A similar bias appears also in RC3 when limiting to
galaxies brighter than 15.5 mag.
This bias might have
important implications while studying the frequencies of bars at low
and high redshifts, especially because the bias
increases toward fainter galaxies.
Also, at high inclinations the number of SAB galaxies in RC3
drops much more rapidly than the number of SB-galaxies, thus
being a manifestation of an ambiguous definition of the de Vaucouleurs's 
SAB class.

\section*{Acknowledgments}
 
We are grateful to Ron Buta for providing us with his H-band image of NGC
1433.
This publication utilized images from the Two Micron All
Sky Survey, which is a joint project of the University of
Massachusetts and the Infrared Processing and Analysis
Center/California Institute of Technology, funded by the National
Aeronautics and Space Administration and the National Science
Foundation. It also uses the NASA/IPAC 
Extra-galactic Database (NED), operated by the Jet Propulsion
Laboratory in Caltech. 
We acknowledge the foundations of Magnus Ehrnrooth and the Academy of 
Finland of significant financial support.

\vfill
\eject

\vfill
\eject

\section*{Figure captions}

{\bf Figure 1.} Different models for the vertical density profile are compared.
In the upper row the commonly used exponential, $sech^2$ and $sech$
laws are compared, with the scale-factor $h$ chosen in a manner that
yields identical slopes for $z \ >> \ h_z$. The frame in the left
displays the vertical profiles, while in the right the resulting
convolution functions $g$ are shown. These models correspond to those
applied in Table 4 for calculation of $Q_b$. The lower row compares
5 different model profiles, including also Gaussian and uniform
models. The scale-factors are chosen in a manner that yields the same
vertical dispersion in each case. For comparison, the convolution
function corresponding to 2D softened gravity is also shown, with softening
parameter $\epsilon=\sqrt{<z^2>}$. These same models are applied for
calculation of $Q_T$-profiles for NGC 1433 in Fig. 3.

{\bf Figure 2.}  Testing of the calculation method. The left hand
frames show the results of our polar method applied to high quality
H-band image of NGC 1433 (Buta et al. 2001), using different maximal numbers
of even Fourier components as indicated in the frame. Fourier
components were calculated with two pixel wide zones (1 pix =1.141
arcsec), and the image was divided to 128 azimuthal bins. In the upper frame
$h_r/h_z=2.5$, typical for early-type galaxies, while the
lower frame shows the influence of assuming 4 times thinner disk (For
$h_r=45''$ and distance of 11.6 Mpc the studied $h_z$ values are about
1kpc and 0.25 kpc, respectively). For comparison, the right-hand
frames show results using a Cartesian evaluation of the
gravitational potential, with different number of grid divisions. The
results for $512 \times 512$ Cartesian grids are also super-posed on the
left-hand curves.

{\bf Figure 3.} The effect of using different vertical density laws on
the obtained $Q_T$-profiles for NGC 1433. The density laws correspond
to those in the lower row of Fig. 1 (scale-factors chosen to yield the
same vertical dispersion), and the plot indicates how progressively
more peaked vertical density profiles yield stronger non-axisymmetric
forces.  The two studied values of $h_r/h_z$ are the same as in Fig.
2.: for thinner disk the precise form of the density profile is less
significant.

{\bf Figure 4.}
The effect of radius-dependent scale-height on $Q_T(r)$ profiles for
NGC 1433.  As in Fig 2, an exponential vertical profile is assumed
for each distance, but with either positive ($dh_z/dr=0.05$, solid
curves) or negative ($dh_z/dr=-0.05$, dashed curves) gradient of the
vertical scale height with distance. Three different cases are
studied, where $h_r/h_z = 2.5$, either at $r= 1, 2$, or $3 h_r$. 
The lines indicate the
vertical scale height-profile in each case, with $h_z$ multiplied by
0.01. According to Schwarzkopf $\&$ Dettmar (2001), the disk 
thickness often increases with
radius, and $dh_z/dr=0.05$ corresponds to the maximum value observed
for early types. Negative gradients are also studied, to emphasize the
smallness of the expected maximal effect of vertically dependent $h_z$.
For comparison, the crosses indicate the peak of the $Q_T$ profile
in the case of constant $h_r/h_z=2.5$.
The polar method is used.

{\bf Figure 5.}  A more detailed comparison of the effects of 
distance-dependent $h_z$,
corresponding to the middle frame in Fig. 4, except that twice larger
positive and negative gradients are studied, together with the case of
constant $h_z$. The mean radial force profile is shown, together with
the $m=2$ and $m=4$ Fourier amplitudes of the tangential force
components (forces are in arbitrary units). As expected, tangential
force components are increased in regions where scale height is
reduced. The same is true also for the radial force, which however
is even more strongly affected in the region of maximal
$F_T/F_R$ ratio, explaining the slightly reduced $Q_T$ in the case of
positive $dh_z/dr$ gradient.

{\bf Figure 6.} Bar strength measurements by BB and by us are compared 
for 13 galaxies in H-band. In both measurements the same orientation parameters
were used and the vertical mass distribution was approximated by
an exponential profile with $h_z=325 \ kpc$ ($H_0 = 75 \ km \ s^{-1} \ Mpc^{-1}$).
The $m=0,2,4,6$
Fourier components of density were included, and calculated for radial 
annulae with width
of $2"$. In the azimuthal direction 128 divisions were used in all cases.
No softening was used in our potential evaluation. 

{\bf Figure 7.} For the ``classical barred'' galaxies in the near-IR
we show: the original 2MASS image in the plane of the sky ($1^{th}$ column),
the same image de-projected to face-on orientation, with the 
m=0 Fourier component being subtracted ($2^{th}$ column), the ``butterfly 
pattern'' ($3^{th}$ column), the m=0,2,4 surface density profiles 
($4^{th}$ column)
and the radial $Q_T$-profile ($5^{th}$ column). The ``butterfly patterns'' were
calculated from the potentials using all even and odd components up to m=6,
while in the $Q_T$-profiles all even components up to 10 were included. 
In the butterfly diagrams the thick contours correspond to $Q_T$
levels of 0.1, 0.2, 0.3, 0.4, 0.5, while the thin contours refer to corresponding
negative levels.
In the $Q_T$-profiles the thin lines show measurements in the four image
quadrants, whereas the thick lines show the mean values. The dashed vertical 
line in the $Q_T$-profile shows the length of the bar estimated from the 
phases of m=2 and m=4 density components
as explained in the text. The numbers in the upper right corner give the values
for the maximum $Q_T$ in the bar region and its radial distance.
The maximum is also shown by the box symbol in the $Q_T$-profiles.
In the de-projected image the solid circle indicates the location
of the maximum $Q_T$ and the dashed circle the measurement region.

{\bf Figure 8.} Absolute blue magnitude v.s. bar length. 
The magnitudes are from RC3 and bar lengths are estimated from the phases
of the m=2 Fourier amplitudes as explained in the text. In Fig. 8a bar 
lengths are given in absolute units, using the distances from
Tully (1988), and in Fig. 8b they are scaled to the scale-length of the
disk. H = 75 $km \ s^{-1} Mpc^{-1} $.

{\bf Figure 9.} The inclination distribution of the galaxies
in our sample, 
SB galaxies and the ``classical bars'' identified in the
near-IR (see the text) being shown separately. For comparison, the 
inclination distribution
for all galaxies in our sample is also shown. The dashed line indicates 
a random distribution of inclinations,
normalized to the maximum inclination and the number of galaxies
in the sample.

{\bf Figure 10.} A correlation between the inclination of the disk (INC)
and bar strength ($Q_b$) for the galaxies in our sample. 

{\bf Figure 11.} A similar figure as Fig. 9, but shown for the de 
Vaucouleurs's (1963) classes SA, SAB and SB 
for a magnitude-limited sample ($B_T < 15.5 \ mag$) from RC3.

{\bf Figure 12.} Distribution of the non-axisymmetric forces among the
de Vaucouleurs's (1963) classes SA, SAB and SB, measured in J, H and K-bands
(Figs a, b and c, respectively).
For comparison, bar strength
distribution is shown also for the ``classical bars'' identified in the 
near-IR. Notice that not all non-axisymmetric forces in the figure are 
associated to bar-like potentials, being due to spiral
arms for some SA and SAB-galaxies.

\vfill
\eject

\begintable*{1}
\caption{{\bf Table 1.} General properties of the galaxies in the sample.}
\halign{%
\rm#\hfil&\qquad\rm#\hfil&\qquad\rm\hfil#&\qquad\rm\hfil
#&\qquad\rm\hfil#&\qquad\rm\hfil#&\qquad\rm#\hfil
&\qquad\rm\hfil#&\qquad\rm#\hfil&\qquad\hfil\rm#\cr
pgc     & ngc   &$T_m$   &$B_T$   & PA    &INC    & Dist.[Mpc]&$h_R$[``] &act& \cr
\noalign{\vskip 10pt}

2081    &N157   &.SXT4.. &11.00   & 40    &49.8   & 20.9  & 35.8(V)(6)  &    & \cr
2437    &N210   &.SXS3.. &11.60   & 160   &48.6   & 20.3  & 15*         &    & \cr 
3051    &N278   &.SXT3.. &11.47   & -     &17.3   & 11.8  & 13*         &    &  \cr
3089    &N289   &.SBT4.. &11.72   & 130   & 44.9  & 19.4  & 14.0(V)(6)  &    & \cr
5619    &N578   &.SXT5.. &11.44   & 110   & 50.9  & 19.5  & 38*         &    &  \cr
5818    &N598   &.SAS6.. &6.27    & 23    & 53.9  & 0.7   &  533.3(V)(6)& HII& \cr
7525    &N772   &.SAS3.. &11.09   & 125(1)&50.2(1)& 32.3  & 56*         &    & \cr
9057    &N908   &.SAS5.. &10.83   & 75    &64.1   & 17.8  & 45*         &    & \cr
10122   &N1042  &.SXT6.. &11.56   &155**  & 12.2  & 16.7  & 57.9(V)(6)  &    & \cr
10266   &N1068  &RSAT3.. &9.61    &70     & 31.7  & 14.4  &  21.4*      &Sy1/Sy2& \cr
10464   &N1084  &.SAS5.. &11.31   &25**   & 55.8  & 17.1  & 19*         &    & \cr
10488   &N1097  &.SBS3.. &10.23   & 130   & 47.5  & 14.5  & 39*         &Sy1 & \cr
10496   &N1087  &.SXT5.. &11.46   & 5     & 52.9  & 19.0  & 20.1(V)(6)  &    & \cr
11479   &N1187  &.SBR5.. &11.34   & 130   &  42.2 & 16.3  & 23.3(V)(6)  &    & \cr
11819   &N1232  &.SXT5.. &10.52   & 108   &29.4   & 21.1  & 42*         &    & \cr
12007   &N1255  &.SXT4.. &11.40   & 117   &  50.9 & 19.9  &  29.8(V)(6) &    & \cr
12412   &N1300  &.SBT4.. &11.11   & 106   &  48.6 & 18.8  & 76.6(V)(6)  &    & \cr
12431   &N1302  &RSBR0.. &11.60   & -     &  17.3 & 20.0  & 65*         &    & \cr
12626   &N1309  &.SAS4*. &11.97   & 45    &  21.0 & 26.0  & 14*         &    & \cr
13059   &N1350  &PSBR2.. &11.16   & 0     &57.5   & 16.9  & 31*         & Sy & \cr
13255   &N1367  &.SXT1.. &11.57   & 135   &  46.2 & 17.1  & 50.6(V)(6)  &    & \cr
13368   &N1385  &.SBS6.. &11.45   & 165   &55.0(5)& 17.5  & 29.2(V)(6)  &    & \cr
13434   &N1398  &PSBR2.. &10.57   & 100   &  40.7 & 16.1  & 29.4(V)(6)  &Sy  & \cr
13602   &N1425  &PSBR2.. &11.29   & 129   &  63.5 & 17.4  & 46.5(V)(6)  &Sy  & \cr
14814   &N1559  &.SBS6.. &11.00   & 64    &60.0   & 14.3  &  31*        &    & \cr
15821   &N1637  &.SXT5.. &11.47   & 15    &  35.6 & 8.9   & 38.0*       &    & \cr
16709   &N1792  &.SAT4.. &10.87   & 137   &59.9   & 13.6  & 34*         &    & \cr
16779   &N1808  &RSXS1.. &10.74   & 133   &53     & 10.8  & 26.9(I)(4)  &Sy2 & \cr 
16906   &N1832  &.SBR4.. &11.96   & 10    &  48.6 & 23.5  & 15.3(V)(6)  &    & \cr
18258   &N2139  &.SXT6.. &11.99   & 0**   & 0**   & 22.4  & 19.7(V)(6)  &    & \cr
18602   &N2196  &PSAS1.. &11.82   & 35    &  39.1 & 28.8  & 27.2(V)(6)  &    & \cr
19531   &N2280  &.SAS6.. &10.90   & 163   &60.7   & 23.2  & 45.1(V)(6)  &    & \cr 
25069   &N2655  &.SXS0.. &10.96   & -     &33.7   & 24.4  & 52*         & Sy2& \cr
26259   &N2835  &.SBT5.. &11.01   & 8     &48.6   & 10.8  & 62.6(V)(6)  &    & \cr
26512   &N2841  &.SAR3*. &10.09   & 147   &64.1   & 12.0  & 45.0(V)(6)  &Sy1/L& \cr
27077   &N2903  &.SXT4.. &9.68    & 17    &60.0(4)& 6.3   & 54.5(V)(6)  &HII & \cr
27777   &N2964  &.SXR4*. &11.99   & 97    &56.7   & 25.9  & 21.2(V)(6)  &HII & \cr
28120   &N2976  &.SA.5P. &10.82   & 141(1)&55.2(1)& 2.1   & 45*         &HII & \cr
28316   &N2985  &PSAT2.. &11.18   & 2(1)  &36.9(1)& 22.4  & 35*         &L   & \cr
28630   &N3031  &.SAS2.. &7.89    & 157   &58.3   & 1.4   & 157.5(V)(6) &Sy1.8/L& \cr
29146   &N3077  &.I.0.P. &10.61   & 45    &33.7   & 2.1   & 44*         &HII & \cr
30087   &N3184  &.SXT6.. &10.36   & 135   &21.1   & 8.7   & 74.0(V)(6)  &HII & \cr
30445   &N3227  &.I.0.P. &11.10   & 155   &47.5   & 23.4  & 23.6(V)(6)  & HII& \cr
31650   &N3310  &.SXR4P. &11.15   & -     &39.1   & 18.7  & 123.0(V)(6) &HII & \cr
31883   &N3338  &.SAS5.. &11.64   & 100   &  51.9 & 22.8  & 25.3(V)(6)  &    & \cr
32183   &N3359  &.SBT5.. &11.03   & 170   &  52.9 & 19.2  & 36*         &HII & \cr
33166   &N3486  &.SXR5.. &11.05   & 111(1)&24.5(1)& 7.4   & 36*         &Sy2 & \cr
33371   &N3504  &RSXS2.. &11.82   & -     &  39.1 & 26.5  & 21.9(V)(6)  &HII & \cr
33390   &N3507  &.SBS3.. &11.73   & 90(2) &21.6(2)& 19.8  & 25.0(K)(5)  &L   & \cr
33550   &N3521  &.SXT4.. &9.83    & 163   &57.0(5)& 7.2   & 37.7(I)(4)  &L   & \cr
34298   &N3596  &.SXT5.  &11.95   & 139(2)& 28.4(2)& 23.0 & 27.9(B)(5)  &    & \cr
35164    &N3675   &.SAS3..&11.00 & 2(1)   & 57.3(1)& 12.8 & 34*         &    & \cr
35193    &N3681   &.SXR4..&11.90 & 103(2) & 32.9(2)& 24.2 & 14.0(K)(5)  &    & \cr
35268    &N3686   &.SBS4..&11.89 & 13.1(2)& 41.4(2)& 23.5 & 27.6(B)(5)  &    & \cr
35616    &N3718   &.SBS1P.&11.59 & 173(1) & 52.0(1)& 17.0 & 41.4(V)(6)  &Sy1 & \cr
35676    &N3726   &.SXR5..&10.91 & 10.0   & 46.4   & 17.0 &  -          &    & \cr
&&&&&&&&& \cr
}
\endtable

\vfill
\eject
\begintable*{1}
\caption{{\bf Table 1.} continued}
\halign{%
\rm#\hfil&\qquad\rm#\hfil&\qquad\rm\hfil#&\qquad\rm\hfil#&\qquad\rm\hfil#&\qquad\rm\hfil#&\qquad\rm#\hfil
&\qquad\rm\hfil#&\qquad\rm#\hfil&\qquad\hfil\rm#\cr
pgc & ngc &$T_m$ &$B_T$ & PA &INC  & Dist.[Mpc]  &$h_R$[``] &act& \cr
\noalign{\vskip 10pt}

36921    &N3898   &.SAS2..&11.60 & 107    & 53.9    & 21.9  & 37.8(V)(6)      &HII/L& \cr
37773    &N4027   &.SBS8..&11.66 & 167    & 40.7    & 25.6  & 25.3(V)(6)      &     & \cr
37999    &N4041   &.SAT4*.&11.88 & -      & 21.1    & 22.7  & 17*             &     & \cr
38150    &N4062   &.SAS5..&11.90 & 100(1) & 65.8(1) & 9.7   & 24.7(V)(6)      &HII  & \cr
38739    &N4151   &PSXT2*.&11.50 & 50(7)  & 16.3(9) & 15.5  & 36*             &Sy1.5& \cr
39578    &N4254   &.SAS5. &10.44 & 40**   & 29.4    & 16.8  & 34.8(V)(6)      &     & \cr
39724    &N4274   &RSBR2..&11.34 & 102(1) & 67.0(1) & 9.7   & 56.9(V)(6)      &     & \cr
39907    &N4293   &RSBS0..&11.26 & 72     & 65.0(3) & 17.0  & 42.9(V)(6)      &L    & \cr
40097    &N4314   &.SBT1..&11.43 & -      & 27.0    & 9.7   & 43.3*           & L   & \cr
40153    &N4321   &.SXS4..&10.05 & 153(2) & 27.1(2) & 16.8  & 64.0(B)(5)      &HII/L& \cr 
40614    &N4394   &RSBR3..&11.73 & -      & 25.0(3) & 16.8  & 33.0(V)(6)      &L    & \cr
40692    &N4414   &.SAT5. &10.96 & 161(1) & 45.6(1) & 9.7   & 27.6(V)(6)      &     & \cr
41024    &N4450   &.SAS2..&10.90 & 171(2) & 46.4(2) & 16.8  & 44.8(K)(5)      &L    & \cr
41333    &N4490   &.SBS7P.&10.22 & 125    & 60.7    & 9.3   & 38.4(V)(6)      &     & \cr
41517    &N4501   &.SAT3..&10.36 & 140    & 57.5    & 16.8  & 47.0(V)(6)      &Sy2  &  \cr
41934    &N4548   &.SBT3..&10.96 & 150    & 37.4    & 16.8  & 28.7(V)(6)      &Sy/L & \cr
42089    &N4569   &.SXT2..&10.26 & 23     & 62.8    & 16.8  & 73.5(V)(6)      & Sy/L& \cr
42575    &N4618   &.SBT9..&11.22 & 25     & 35.6    & 8.2   & 60*             &HII  & \cr
42833    &N4651   &.SAT5..&11.39 & 73.9(2)& 47.9(2) &16.8   & 25.1(B)(5)      &L    & \cr
42857    &N4654   &.SXT6..&11.10 & 128    & 57.0(3) & 16.8  & 39.8(V)(6)      &     & \cr
43186    &N4689   &.SAT4..&11.60 & 155**  & 40.0(3) & 16.8  & 38.9(V)(6)      &     & \cr
43238    &N4691   &RSBS0P.&11.66 & 15     & 35.6(4) & 22.5  & 35.8(V)(6)      &     & \cr
43321    &N4699   &.SXT3..&10.41 & 45     & 46.2    & 25.7  & -               &Sy   & \cr
43495    &N4736   &RSAR2..&8.99  & 105    & 35.6    & 4.3   & 84.3(V)(6)      &L    & \cr
43507    &N4731   &.SBS6..&11.90 & 85     & 60.7    & 25.9  & 39*             &     & \cr
43671    &N4753   &.I.0...&10.85 & 80     & 62.1    & 15.1  & 46*             &     & \cr
44182    &N4826   &RSAT2..&9.36  & 115    & 57.     & 4.1   & 61.9(V)(6)      & Sy2 & \cr
45165    &N4941   &RSXR2*.&12.43 & 15     & 57.5    & 6.4   & 21.6(V)(6)      &Sy2  & \cr
45749    &N5005   &.SXT4..&10.61 & 67(1)  & 65.0**  & 21.3  & 33.0(V)(6)      &Sy2  & \cr
45948    &N5033   &.SAS5..&10.75 & 173(1) & 60.0(1) & 18.7  & 731.6(V)(6)     &Sy1.9&  \cr
46247    &N5054   &.SAS4..&11.67 & 155    & 54.0(4) & 27.3  & 40.3(I)(4)      &     & \cr
46400    &N5068   &.SXT6..&10.52 & 110    & 29.4    & 6.7   & 39.3(V)(6)      &     & \cr
47404    &N5194   &.SAS4P.&8.96  & 170(8) & 20.0(8) & 9.3   & 107.5(V)(6)     &HII/Sy2.5& \cr
47413    &N5195   &.I.0.P.&10.45 & 90(9)  & 30.0(9) & 7.7   & 29*             &L    & \cr
48171    &N5247   &.SAS4..&10.50 & 20     & 29.4    & 22.2  & 62*             &     & \cr
50063    &N5457   &.SXT6..&8.31 & -       & 21.1    & 5.4   & 143.4(V)(6)     &     & \cr
52412    &N5713   &.SXT4P.&12.18 & 10     & 27.0    & 30.4  & 13.8(V)(6)      &     & \cr
55588    &N5962   &.SAR5..&11.98 & 96.1(2)& 44.8 (2)& 31.8  & 17.9(B)(5)      &HII  & \cr
56219    &N6015   &.SAS6..&11.69 & 28(1)  & 61.3(1) & 17.5  & 37.1(V)(6)      &HII  & \cr
58477    &N6217   &RSBT4. &11.79 & 150**  & 33.7    & 23.9  & 13.2(V)(6)      &Sy2  & \cr
61742    &N6643   &.SAT5..&11.73 & 40(1)  & 60.0(1) & 25.5  & 27.7(V)(6)      &     & \cr
68096    &N7217   &RSAR2..&11.02 & 95     & 33.7    & 16.0  & 26.9(V)(6)      &Sy   & \cr
68165    &N7213   &.SAS1*.&11.01 & -      & 26.9    & 22.0  & 82*             &Sy1.5& \cr
69253    &N7314   &.SXT4..&11.88 & 3      & 62.8    & 18.3  & -               &Sy1.9& \cr
69327    &N7331   &.SAS3..&10.35 & 170(1) & 62.6(1) & 14.3  & 89.4(V)(6)      &L    & \cr 
70419    &N7479   &.SBS5..&11.60 & 40.5(2)& 39.6(2) &32.4   & 50.7(B)(5)      &Sy2/L& \cr
70714    &N7513   &PSBS3P.&11.39 &108     & 48.6    &21.3   & 55*             &     & \cr
71047    &N7606   &.SAS3..&11.51 & 146(1) & 67.7(1) & 28.9  & 43*             &     & \cr
72009    &N7723   &.SBR3..&11.94 & 35     & 47.5    & 23.7  & 19*             &     & \cr
72060    &N7727   &.SXS1P.&11.50 & 35     & 40.7    & 23.3  &  -              &     & \cr
72237    &N7741   &.SBS6..&11.84 & 170(2) & 45.6(2) & 12.3  & 52.7(B)(5)      &     & \cr
}
\tabletext{\noindent (1) Heraudeau \& Simien (1996).

\noindent (2) de Jong \& van der Kruit (1994). 



\noindent (3) Yasuda et al. (1995).

\noindent (4) Roth (1994).

\noindent (5) de Jong 1996.

\noindent (6) Bagget et al. 1998.

\noindent (7) Sanchez-Portal et al. (2000).

\noindent (8) Tully (1974).

\noindent (9) Smith et al. (1990).

\noindent * estimated from 2MASS images.

\noindent ** own estimate from the Digitized Sky Survey plate.

}
\endtable

\vfill
\eject

\begintable*{1}
\caption{{\bf Table 2.} $Q_b$, $r_{Qb}$, $\epsilon$ and $r_{\epsilon}$ for 
galaxies with classical bars in the near-IR.}
\halign{%
\rm#\hfil&\qquad\rm#\hfil&\qquad\rm\hfil#&\qquad\rm\hfil
#&\qquad\rm\hfil#&\qquad\rm\hfil#&\qquad\rm#\hfil
&\qquad\rm\hfil#&\qquad\rm#\hfil&\qquad\hfil\rm#\cr
pgc & $Q_H$  & $r_H$[``] & $Q_J$ &$r_J$[``]  & $Q_K$ &$r_K$[``]  &length [``] &$\epsilon$ &$r_{\epsilon}$[``] \cr
\noalign{\vskip 10pt}

3089      &0.15 $\pm$ 0.00 &13 &0.14 $\pm$ 0.00 &15 &0.17 $\pm$ 0.03 & 15 &20  &0.39  &    20.28 \cr
10266     &0.15 $\pm$ 0.01 &11 &0.16 $\pm$ 0.02 &11 &0.12 $\pm$ 0.01 & 11 &25  &0.43  &    14.96 \cr
10488     &0.24 $\pm$ 0.01 &79 &0.24 $\pm$ 0.02 &79 &0.23 $\pm$ 0.01 & 77 &120 &0.58  &    94.72 \cr
10496     &0.42 $\pm$ 0.00 &11 &0.43 $\pm$ 0.01 &11 &0.44 $\pm$ 0.01 & 11 &22  &0.63  &    24.26 \cr
11479     &0.19 $\pm$ 0.03 &29 &0.19 $\pm$ 0.02 &31 &0.21 $\pm$ 0.03 & 27 &40  &0.55  &    41.42 \cr
12412     &0.40 $\pm$ 0.03 &61 &0.38 $\pm$ 0.01 &59 &0.41 $\pm$ 0.04 & 59 &80  &0.71  &    68.77 \cr
12431     &0.09 $\pm$ 0.02 &19 &0.09 $\pm$ 0.01 &19 &0.09 $\pm$ 0.04 & 19 &30  &0.35  &    28.06 \cr
13059     &0.29 $\pm$ 0.01 &69 &0.30 $\pm$ 0.00 &71 &0.30 $\pm$ 0.00 & 69 &110 &0.62  &    77.30 \cr
13368     &0.37 $\pm$ 0.01 &35 &0.35 $\pm$ 0.02 &35 &0.35 $\pm$ 0.02 & 11 &45  &0.70  &     9.91 \cr
13434     &0.20 $\pm$ 0.00 &39 &0.21 $\pm$ 0.01 &39 &0.20 $\pm$ 0.00 & 39 &60  &0.54  &    49.76 \cr
13602     &0.16 $\pm$ 0.00 &11 &0.15 $\pm$ 0.01 &11 &0.15 $\pm$ 0.01 & 11 &25  &0.48  &     4.73 \cr
14814     &0.30 $\pm$ 0.02 &11 &0.27 $\pm$ 0.02 &13 &0.24 $\pm$ 0.00 & 13 &15  &0.51  &    13.58 \cr
15821     &0.20 $\pm$ 0.05 &19 &0.23 $\pm$ 0.00 &18 &0.24 $\pm$ 0.04 & 17 &25  &0.56  &    24.74 \cr
16906     &0.23 $\pm$ 0.01 &13 &0.23 $\pm$ 0.00 &15 &0.22 $\pm$ 0.01 & 13 &20  &0.49  &    18.46 \cr
18258     &0.35 $\pm$ 0.01 &7  &0.33 $\pm$ 0.01 &5  &0.35 $\pm$ 0.07 & 7  &25  &0.56  &     0.00 \cr
26259     &0.28 $\pm$ 0.01 &15 &0.31 $\pm$ 0.00 &11 &0.38 $\pm$ 0.01 & 15 &30  &0.58  &    33.18 \cr
27077     &0.26 $\pm$ 0.05 &61 &0.25 $\pm$ 0.01 &63 &0.27 $\pm$ 0.00 & 61 &75  &0.65  &    65.81 \cr
27777     &0.32 $\pm$ 0.00 &21 &0.33 $\pm$ 0.01 &21 &0.33 $\pm$ 0.01 & 23 &30  &0.63  &    26.31 \cr
32183     &0.46 $\pm$ 0.05 &11 &0.42 $\pm$ 0.01 &11 &0.45 $\pm$ 0.03 & 11 &55  &0.65  &    9.46 \cr
33371     &0.26 $\pm$ 0.02 &27 &0.27 $\pm$ 0.03 &29 &0.26 $\pm$ 0.02 & 29 &45  &0.63  &    30.91 \cr
33390     &0.19 $\pm$ 0.04 &17 &0.21 $\pm$ 0.00 &21 &0.20 $\pm$ 0.06 & 21 &20  &0.44  &    20.99 \cr
35268     &0.33 $\pm$ 0.01 &11 &0.28 $\pm$ 0.02 &13 &0.30 $\pm$ 0.01 & 11 &29  &0.59  &    22.25 \cr
35616     &0.12 $\pm$ 0.01 &65 &0.16 $\pm$ 0.02 &60 &0.14 $\pm$ 0.01 & 60 &80  &0.38  &    50.00 \cr
37773     &0.50 $\pm$ 0.01 &11 &0.51 $\pm$ 0.02 &11 &0.48 $\pm$ 0.02 & 11 &20  &0.67  &     7.58  \cr
37999     &0.13 $\pm$ 0.02 &11 &0.08 $\pm$ 0.00 &11 &0.09 $\pm$ 0.01 & 11 &20  &0.33  &     6.07 \cr
38739     &0.16 $\pm$ 0.02 &67 &0.18 $\pm$ 0.03 &67 &0.14 $\pm$ 0.01 & 67 &100 &0.52  &    75.72 \cr
39724     &0.31 $\pm$ 0.03 &45 &0.31 $\pm$ 0.00 &41 &0.31 $\pm$ 0.02 & 45 &100 &0.65  &    66.93 \cr
39907     &0.22 $\pm$ 0.06 &47 &0.22 $\pm$ 0.01 &49 &0.21 $\pm$ 0.02 & 47 &90  &0.50  &    60.00 \cr
40097     &0.33 $\pm$ 0.01 &47 &0.33 $\pm$ 0.00 &45 &0.34 $\pm$ 0.00 & 45 &88  &0.68  &    67.71 \cr
40614     &0.22 $\pm$ 0.01 &29 &0.21 $\pm$ 0.00 &27 &0.22 $\pm$ 0.01 & 25 &65  &0.61  &    43.26 \cr
41024     &0.13 $\pm$ 0.01 &39 &0.13 $\pm$ 0.01 &39 &0.17 $\pm$ 0.01 & 41 &60  &0.41  &    46.65 \cr
41934     &0.33 $\pm$ 0.02 &49 &0.32 $\pm$ 0.01 &47 &0.32 $\pm$ 0.00 & 45 &70  &0.62  &    76.98 \cr
42575     &0.40 $\pm$ 0.03 &10 &0.36 $\pm$ 0.03 &11 &0.36 $\pm$ 0.01 & 13 &45  &0.65  &    16.72 \cr
43238     &0.61 $\pm$ 0.01 &11 &0.59 $\pm$ 0.01 &13 &0.63 $\pm$ 0.00 & 11 &60  &0.77  &    17.97 \cr
43507     &1.19 $\pm$ 0.10 &23 &1.07 $\pm$ 0.00 &25 &1.27 $\pm$ 0.02 & 25 &100 &0.86  &     0.00 \cr
46400     &0.43 $\pm$ 0.03 &11 &0.41 $\pm$ 0.01 &13 &0.48 $\pm$ 0.04 & 13 &52  &0.65  &    20.00 \cr
47413     &0.14 $\pm$ 0.01 &39 &0.15 $\pm$ 0.00 &39 &0.13 $\pm$ 0.01 & 41 &48  &0.46  &    51.49 \cr
52412     &0.37 $\pm$ 0.04 &11 &0.39 $\pm$ 0.05 &13 &0.39 $\pm$ 0.02 & 13 &28  &0.57  &    20.21 \cr
58477     &0.36 $\pm$ 0.01 &29 &0.40 $\pm$ 0.01 &25 &0.40 $\pm$ 0.04 & 29 &40  &0.69  &    35.98 \cr
70419     &0.63 $\pm$ 0.03 &37 &0.62 $\pm$ 0.02 &39 &0.65 $\pm$ 0.04 & 43 &70  &0.78  &    49.71 \cr
70714     &0.46 $\pm$ 0.02 &17 &0.41 $\pm$ 0.01 &25 &0.46 $\pm$ 0.01 & 20 &55  &0.76  &    37.84 \cr
72009     &0.30 $\pm$ 0.02 &16 &0.28 $\pm$ 0.03 &15 &0.31 $\pm$ 0.01 & 17 &28  &0.64  &    20.20 \cr
72237     &0.64 $\pm$ 0.02 &11 &0.70 $\pm$ 0.02 &16 &0.70 $\pm$ 0.05 & 13 &70  &0.77  &     0.00 \cr
}
\endtable
\vfill
\eject




\begintable*{1}
\caption{{\bf Table 3.}Bar strengths for galaxies without classical bars.}
\halign{%
\rm#\hfil&\qquad\rm#\hfil&\qquad\rm\hfil#&\qquad\rm\hfil
#&\qquad\rm\hfil#&\qquad\rm\hfil#&\qquad\rm#\hfil
&\qquad\rm\hfil#&\qquad\rm#\hfil&\qquad\hfil\rm#\cr
     pgc & $Q_H$   & $r_H$[``] & $Q_J$ &$r_J$[``] & $Q_K$ &$r_K$[``]&comment&& \cr
\noalign{\vskip 10pt}

  2081  &0.29 $\pm$ 0.11 &33   &0.26 $\pm$ 0.00 &31 &0.26 $\pm$ 0.01 & 35 & S && \cr
  2437  &0.12 $\pm$ 0.00 &33   &0.12 $\pm$ 0.01 &33 &0.11 $\pm$ 0.00 & 33 & BM && \cr
  3051  &0.06 $\pm$ 0.02 &23   &0.06 $\pm$ 0.01 &25 &0.06 $\pm$ 0.01 & 23 & && \cr
  5619  &0.32 $\pm$ 0.01 &11   &0.28 $\pm$ 0.01 &11 &0.33 $\pm$ 0.08 & 11 & BM && \cr
  5818  &0.24 $\pm$ 0.00 &49   &0.25 $\pm$ 0.04 &35 &0.29 $\pm$ 0.00 & 35 & BM && \cr
  7525  &0.10 $\pm$ 0.02 &33   &0.10 $\pm$ 0.02 &35 &0.12 $\pm$ 0.00 & 35 & && \cr
  9057  &0.18 $\pm$ 0.02 &71   &0.19 $\pm$ 0.00 &67 &0.17 $\pm$ 0.02 & 61 & S && \cr
 10122  &0.38 $\pm$ 0.00 &25   &0.35 $\pm$ 0.06 &23 &0.38 $\pm$ 0.02 & 25 & BM && \cr
 10464  &0.18 $\pm$ 0.02 &29   &0.18 $\pm$ 0.01 &33 &0.19 $\pm$ 0.03 & 33 & B && \cr
 11819  &0.17 $\pm$ 0.01 &11   &0.16 $\pm$ 0.01 &11 &0.20 $\pm$ 0.03 & 11 & BM && \cr
 12007  &0.15 $\pm$ 0.02 &25   &0.15 $\pm$ 0.02 &20 &0.18 $\pm$ 0.03 & 25 & S && \cr
 12626  &0.15 $\pm$ 0.06 &15   &0.12 $\pm$ 0.02 &15 &0.14 $\pm$ 0.02 & 15 & B && \cr
 13255  &0.12 $\pm$ 0.02 &21   &0.12 $\pm$ 0.00 &21 &0.15 $\pm$ 0.00 & 23 & BM && \cr
 16709  &0.25 $\pm$ 0.00 &33   &0.24 $\pm$ 0.00 &33 &0.26 $\pm$ 0.00 & 33 & S && \cr
 16779  &0.25 $\pm$ 0.00 &69   &0.27 $\pm$ 0.00 &67 &0.23 $\pm$ 0.00 & 71 & BM && \cr
 18602  &0.07 $\pm$ 0.00 &     &0.09 $\pm$ 0.02 &   &0.10 $\pm$ 0.04 &    & && \cr
 19531  &0.17 $\pm$ 0.02 &45   &0.17 $\pm$ 0.02 &45 &0.17 $\pm$ 0.03 & 45 & BM && \cr
 25069  &0.12 $\pm$ 0.01 &     &0.13 $\pm$ 0.01 &   &0.13 $\pm$ 0.03 &    & B && \cr
 26512  &0.18 $\pm$ 0.01 &13   &0.19 $\pm$ 0.01 &13 &0.19 $\pm$ 0.01 & 13 & BM && \cr
 28120  &0.30 $\pm$ 0.01 &57   &0.30 $\pm$ 0.02 &61 &0.35 $\pm$ 0.01 & 63 & BM && \cr
 28316  &0.03 $\pm$ 0.03 &     &0.02 $\pm$ 0.01 &   &0.02 $\pm$ 0.02 &    &  && \cr
 28630  &0.06 $\pm$ 0.02 &     &0.06 $\pm$ 0.02 &   &0.06 $\pm$ 0.02 &    &  && \cr
 29146  &0.07 $\pm$ 0.03 &     &0.09 $\pm$ 0.01 &   &0.07 $\pm$ 0.03 &    &  && \cr
 30087  &0.16 $\pm$ 0.01 &45   &0.15 $\pm$ 0.01 &45 &0.15 $\pm$ 0.01 &45  & S && \cr
 30445  &0.18 $\pm$ 0.07 &49   &0.18 $\pm$ 0.01 &41 &0.19 $\pm$ 0.00 & 53 & B && \cr
 31650  &0.06 $\pm$ 0.01 &     &0.04 $\pm$ 0.00 &   &0.07 $\pm$ 0.00 &    & && \cr
 31883  &0.12 $\pm$ 0.01 &45   &0.14 $\pm$ 0.05 &49 &0.16 $\pm$ 0.03 & 45 & S && \cr
 33166  &0.16 $\pm$ 0.00 &17   &0.15 $\pm$ 0.00 &60 &0.16 $\pm$ 0.00 & 60 & BM && \cr
 33550  &0.02 $\pm$ 0.00 &     &0.06 $\pm$ 0.01 &   &0.06 $\pm$ 0.01 &    &  && \cr
 34298  &0.16 $\pm$ 0.01 &23   &0.16 $\pm$ 0.04 &19 &0.16 $\pm$ 0.02 & 34 & S && \cr
 35164  &0.04 $\pm$ 0.01 &     &0.04 $\pm$ 0.00 &   &0.04 $\pm$ 0.03 &    & S & & \cr
 35193  &0.25 $\pm$ 0.01 & 8   &0.21 $\pm$ 0.02 & 9 &0.23 $\pm$ 0.01 &  8 & BM && \cr
 35676  &0.33 $\pm$ 0.04 &23   &0.30 $\pm$ 0.01 &27 &0.30 $\pm$ 0.05 & 29 & BM && \cr
 36921  &0.03 $\pm$ 0.02 &     &0.02 $\pm$ 0.02 &   &0.02 $\pm$ 0.00 &    &  && \cr
 38150  &0.14 $\pm$ 0.03 &47   &0.09 $\pm$ 0.00 &50 &0.13 $\pm$ 0.02 & 47 & BM && \cr
 39578  &0.12 $\pm$ 0.01 &19   &0.13 $\pm$ 0.01 &23 &0.13 $\pm$ 0.02 & 21 & S && \cr
 40153  &0.20 $\pm$ 0.07 &59   &0.20 $\pm$ 0.02 &59 &0.22 $\pm$ 0.02 & 61 & S && \cr
 40692  &0.11 $\pm$ 0.01 &     &0.11 $\pm$ 0.00 &   &0.11 $\pm$ 0.00 &    & S && \cr
 41333  &0.25 $\pm$ 0.01 &73   &0.25 $\pm$ 0.01 &73 &0.25 $\pm$ 0.01 &73  & BM && \cr
 41517  &0.07 $\pm$ 0.01 &     &0.07 $\pm$ 0.00 &   &0.07 $\pm$ 0.00 &    & && \cr
 42089  &0.15 $\pm$ 0.01 &61   &0.14 $\pm$ 0.01 &61 &0.17 $\pm$ 0.01 & 59 & BM && \cr
 42833  &0.05 $\pm$ 0.01 &17   &0.05 $\pm$ 0.01 &13 &0.12 $\pm$ 0.01 & 19 & BM && \cr
 42857  &0.14 $\pm$ 0.02 &     &0.14 $\pm$ 0.00 &   &0.14 $\pm$ 0.00 &    & S && \cr
 43186  &0.03 $\pm$ 0.01 &     &0.03 $\pm$ 0.00 &   &0.04 $\pm$ 0.02 &    &  && \cr
 43321  &0.04 $\pm$ 0.01 &     &0.04 $\pm$ 0.02 &   &0.04 $\pm$ 0.01 &    &  && \cr
 43495  &0.02 $\pm$ 0.01 &     &0.02 $\pm$ 0.01 &   &0.02 $\pm$ 0.01 &    & && \cr
 43671  &0.21 $\pm$ 0.00 &     &0.20 $\pm$ 0.00 &   &0.20 $\pm$ 0.01 &    &B & & \cr
 44182  &0.13 $\pm$ 0.01 &     &0.12 $\pm$ 0.00 &   &0.13 $\pm$ 0.00 &    & && \cr
 45165  &0.05 $\pm$ 0.00 &     &0.04 $\pm$ 0.02 &   &0.04 $\pm$ 0.05 &    &  && \cr
 45749  &0.16 $\pm$ 0.01 &25   &0.15 $\pm$ 0.01 &25 &0.17 $\pm$ 0.01 &28  & BM && \cr
 45948  &0.07 $\pm$ 0.04 &     &0.04 $\pm$ 0.03 &   &0.05 $\pm$ 0.05 &    &  && \cr
 46247  &0.15 $\pm$ 0.03 &45   &0.16 $\pm$ 0.04 &47 &0.12 $\pm$ 0.05 &45  & S && \cr
 47404  &0.16 $\pm$ 0.00 &113  &0.17 $\pm$ 0.01 &113&0.17 $\pm$ 0.00 &111 & S && \cr
 48171  &0.25 $\pm$ 0.04 &50   &0.27 $\pm$ 0.02 &61 &0.24 $\pm$ 0.05 & 61 & S && \cr
 50063  &0.10 $\pm$ 0.02 &     &0.14 $\pm$ 0.02 &   &0.15 $\pm$ 0.02 &    & S && \cr
 55588  &0.22 $\pm$ 0.03 &13   &0.21 $\pm$ 0.00 &15 &0.26 $\pm$ 0.02 & 15 & BM && \cr
 56219  &0.08 $\pm$ 0.01 &     &0.09 $\pm$ 0.04 &   &0.09 $\pm$ 0.01 &    &  && \cr
 61742  &0.16 $\pm$ 0.10 &     &0.20 $\pm$ 0.01 &   &0.20 $\pm$ 0.01 &    & B && \cr
 68096  &0.05 $\pm$ 0.01 &     &0.06 $\pm$ 0.00 &   &0.05 $\pm$ 0.00 &    &  && \cr
 68165  &0.01 $\pm$ 0.01 &     &0.01 $\pm$ 0.02 &   &0.02 $\pm$ 0.00 &    &  && \cr
 69253  &0.27 $\pm$ 0.04 &58   &0.22 $\pm$ 0.02 &50 &0.22 $\pm$ 0.03 & 50 & BM && \cr
 69327  &0.14 $\pm$ 0.03 &45   &0.10 $\pm$ 0.01 &33 &0.11 $\pm$ 0.01 &53  & S && \cr
 71047  &0.12 $\pm$ 0.02 &     &0.07 $\pm$ 0.01 &   &0.09 $\pm$ 0.00 &    & && \cr
 72060  &0.20 $\pm$ 0.00 &13   &0.23 $\pm$ 0.00 &13 &0.22 $\pm$ 0.00 &13  & BM && \cr
}
\tabletext{\noindent
S = $Q_T$ due to spiral arms
B = bar-like ``butterfly'' structure, but no clear $Q_T$-maximum in the assumed bar region
BM= both $Q_T$-maximum and bar-like butterfly structure appear}
\endtable

\vfill
\eject

\begintable*{1}
\caption{{\bf Table 4.} For the test galaxies: bar strengths and the parameters used in the calculation.}
\halign{%
\rm#\hfil&\qquad\rm#\hfil&\qquad\rm\hfil#&\qquad\rm\hfil
#&\qquad\rm\hfil#&\qquad\rm\hfil#&\qquad\rm#\hfil
&\qquad\rm\hfil#&\qquad\rm#\hfil&\qquad\hfil\rm#\cr

Galaxy &   PA & INC & \ \ DIST [Mpc]  & $Q_b(sech^2)$   & $Q_b(sech)$ & $Q_b(exp)$  & $Q_b(BB)$ && \cr
\noalign{\vskip 10pt}
Maffei2  &  23 &65.1&  3.4  & 0.33 &0.34 &0.35 &0.27 $\pm$ 0.03 && \cr
E565-G11 &  73 &33.0& 63.0  & 0.39 &0.41 &0.42 &0.28 $\pm$ 0.03 && \cr
E566-G24 &  67 &42.5& 45.0  & 0.36 &0.37 &0.39 &0.27 $\pm$ 0.04 && \cr
N309     & 175 &33.7& 75.5  & 0.26 &0.27 &0.28 &0.11 $\pm$ 0.02 && \cr
N1300    & 106 &48.6& 18.8  & 0.55 &0.58 &0.60 &0.42 $\pm$ 0.06 && \cr
N1637    &  15 &35.6&  8.9  & 0.20 &0.21 &0.22 &0.09 $\pm$ 0.03 && \cr
N2543    &  45 &54.9& 32.9  & 0.31 &0.33 &0.35 &0.28 $\pm$ 0.05 && \cr
N3081    & 123 &33.0& 32.5  & 0.28 &0.28 &0.29 &0.17 $\pm$ 0.02 && \cr
N4548    & 150 &37.4& 16.8  & 0.38 &0.39 &0.40 &0.44 $\pm$ 0.03 && \cr
N4653    &  30 &29.4& 39.1  & 0.20 &0.22 &0.23 &0.04 $\pm$ 0.01 && \cr
N5371    &   8 &37.4& 34.1  & 0.19 &0.20 &0.21 &0.19 $\pm$ 0.02 && \cr
N5905    & 135 &48.6& 42.5  & 0.43 &0.44 &0.45 &0.43 $\pm$ 0.05 && \cr
N7479    &  25 &40.6& 32.4  & 0.73 &0.76 &0.79 &0.63 $\pm$ 0.08 && \cr
}
\endtable

\vfill
\eject


\psfig{file=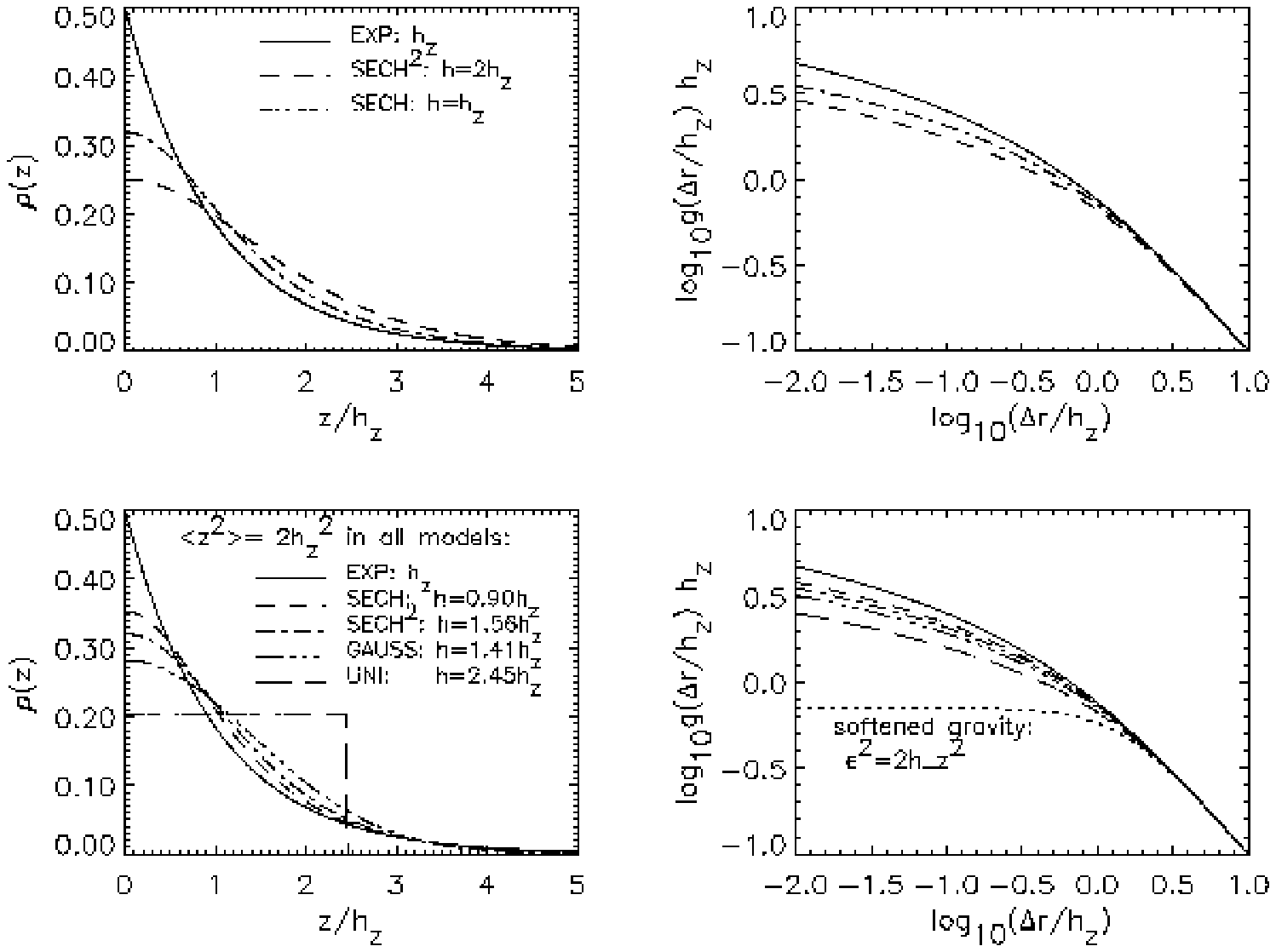,width=16cm}  
Fig. 1
\vfill
\eject

\psfig{file=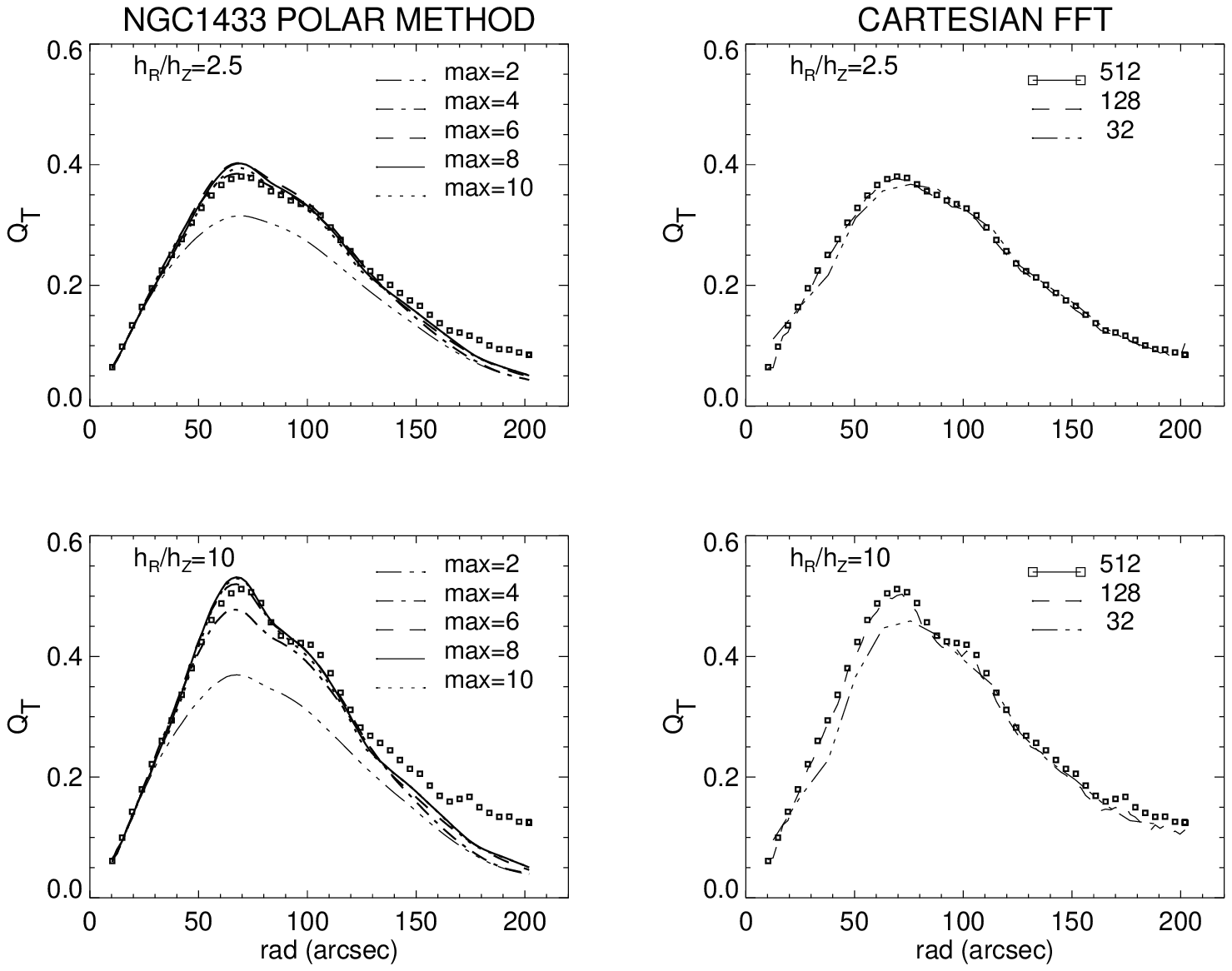,width=16cm}    
Fig. 2
\vfill
\eject

\psfig{file=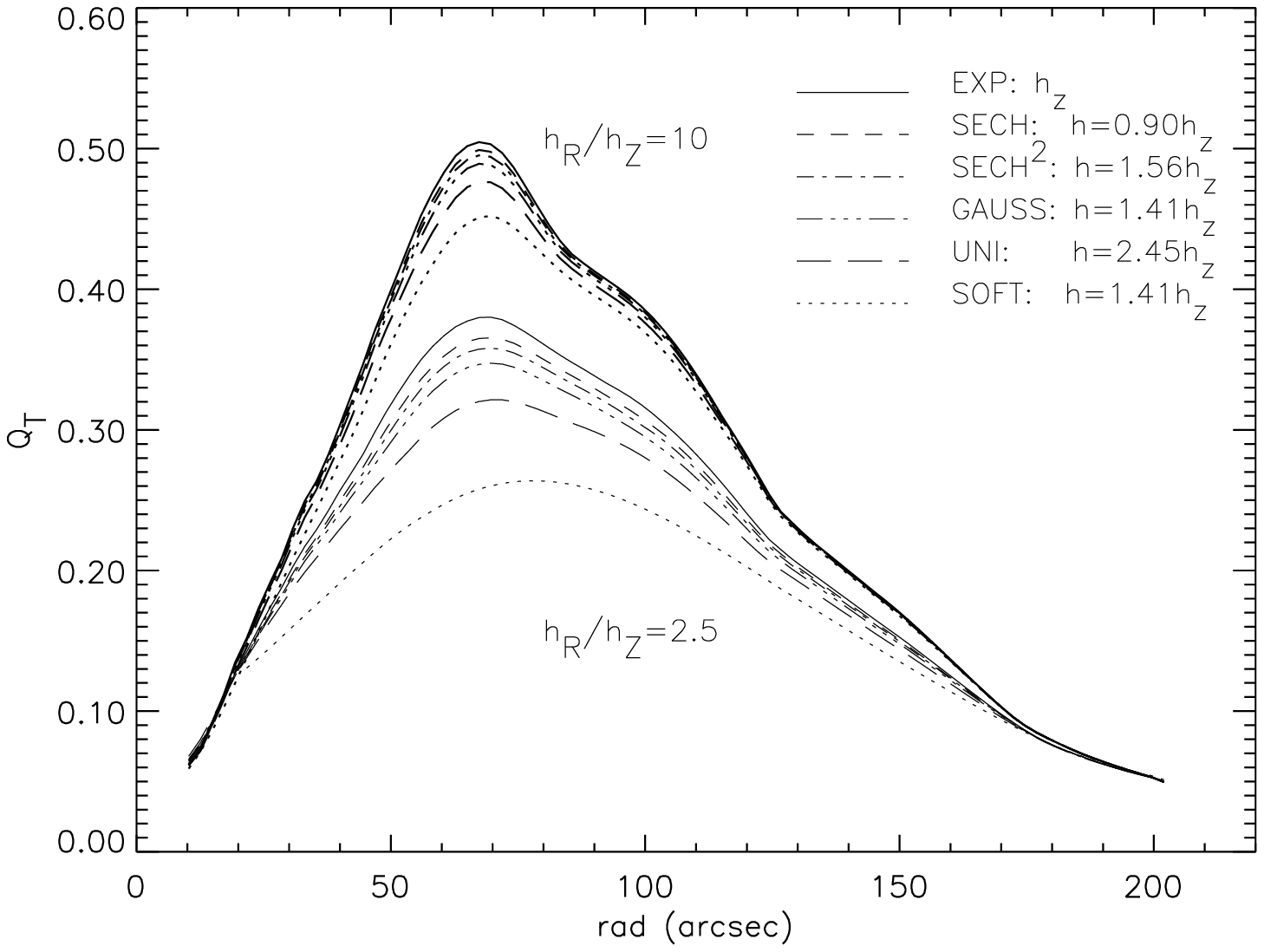,width=16cm} 
Fig. 3
\vfill
\eject

\psfig{file=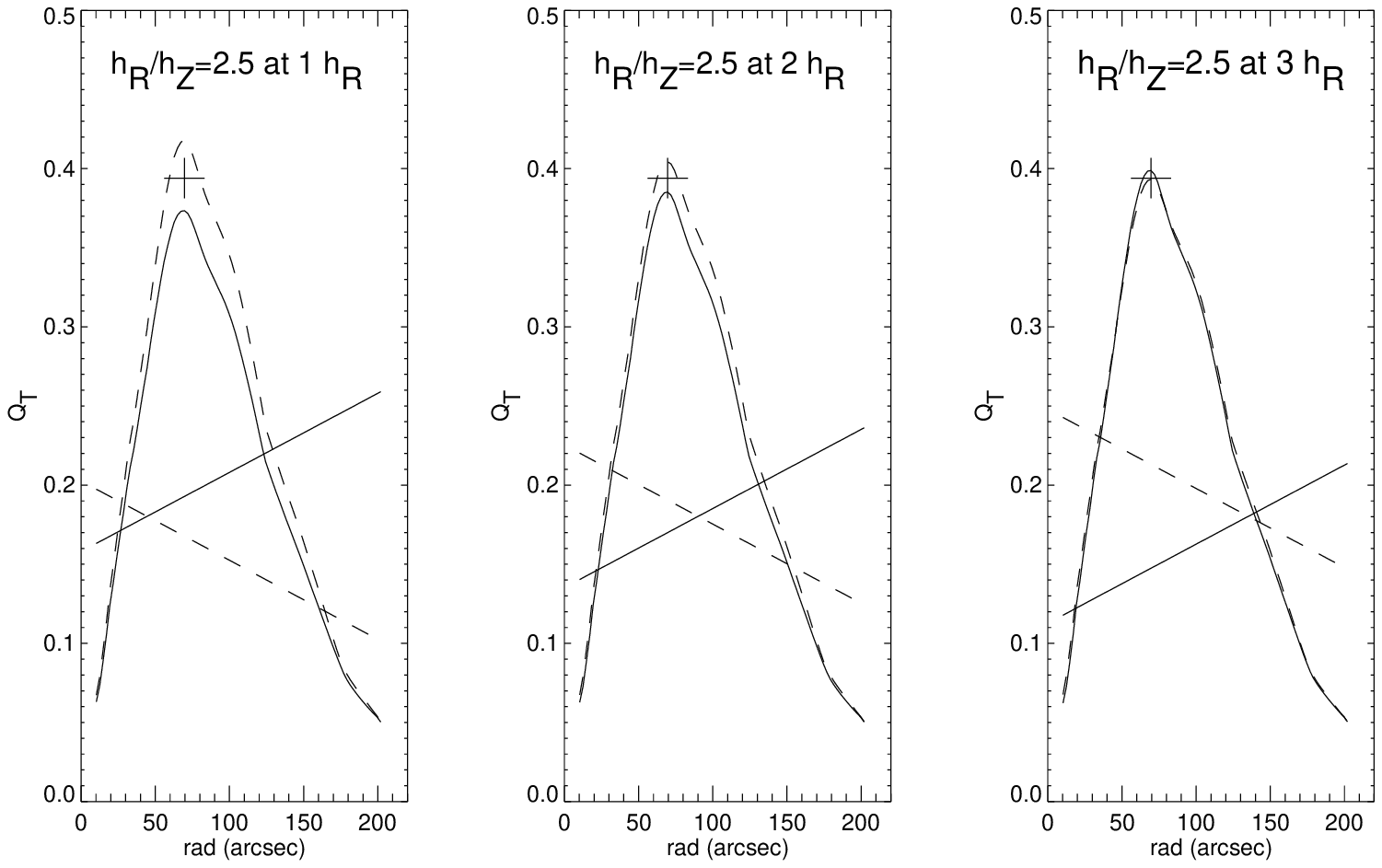,width=16cm} 
Fig. 4
\vfill
\eject

\psfig{file=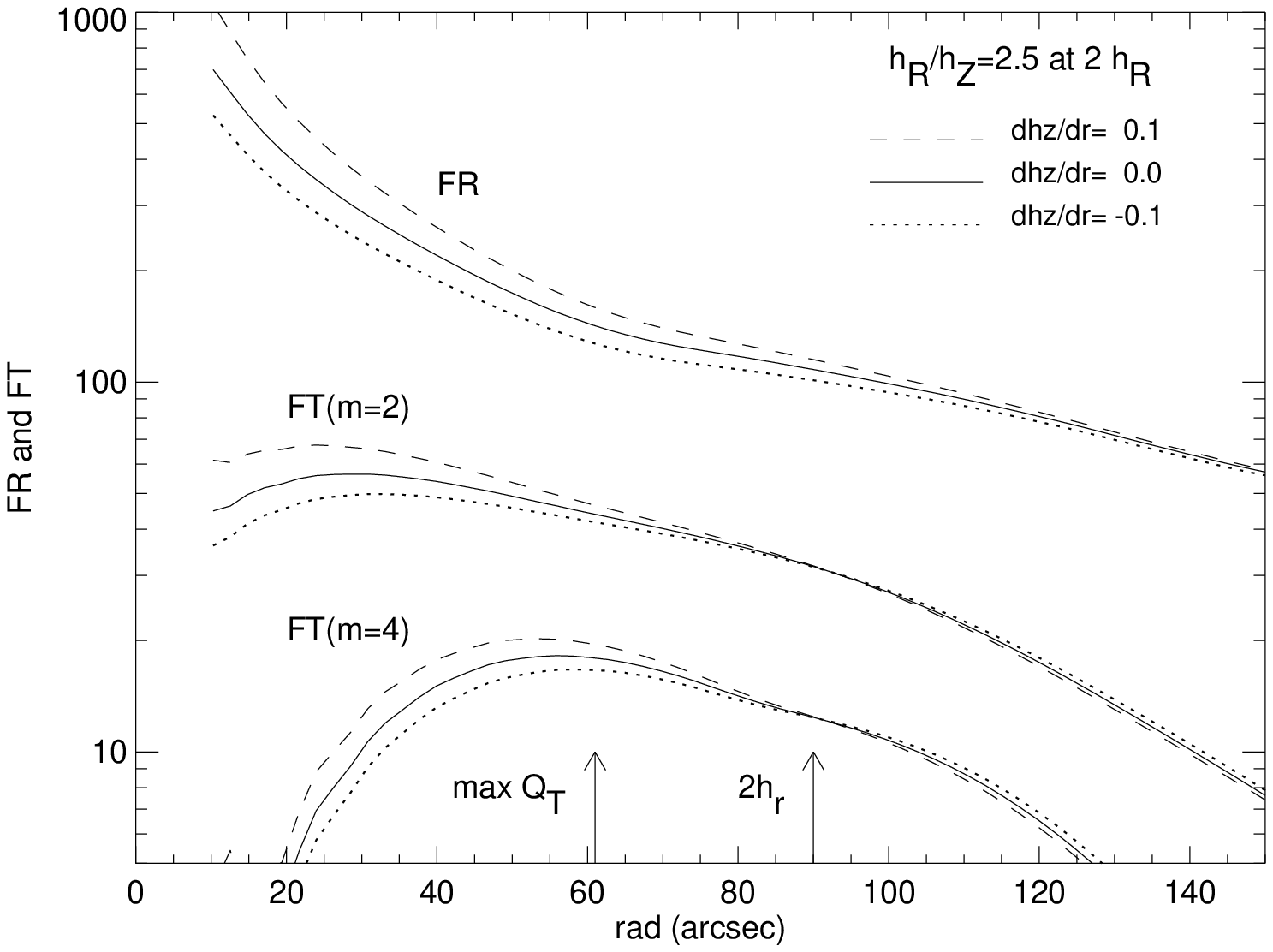,width=16cm} 
Fig. 5
\vfill
\eject

\psfig{file=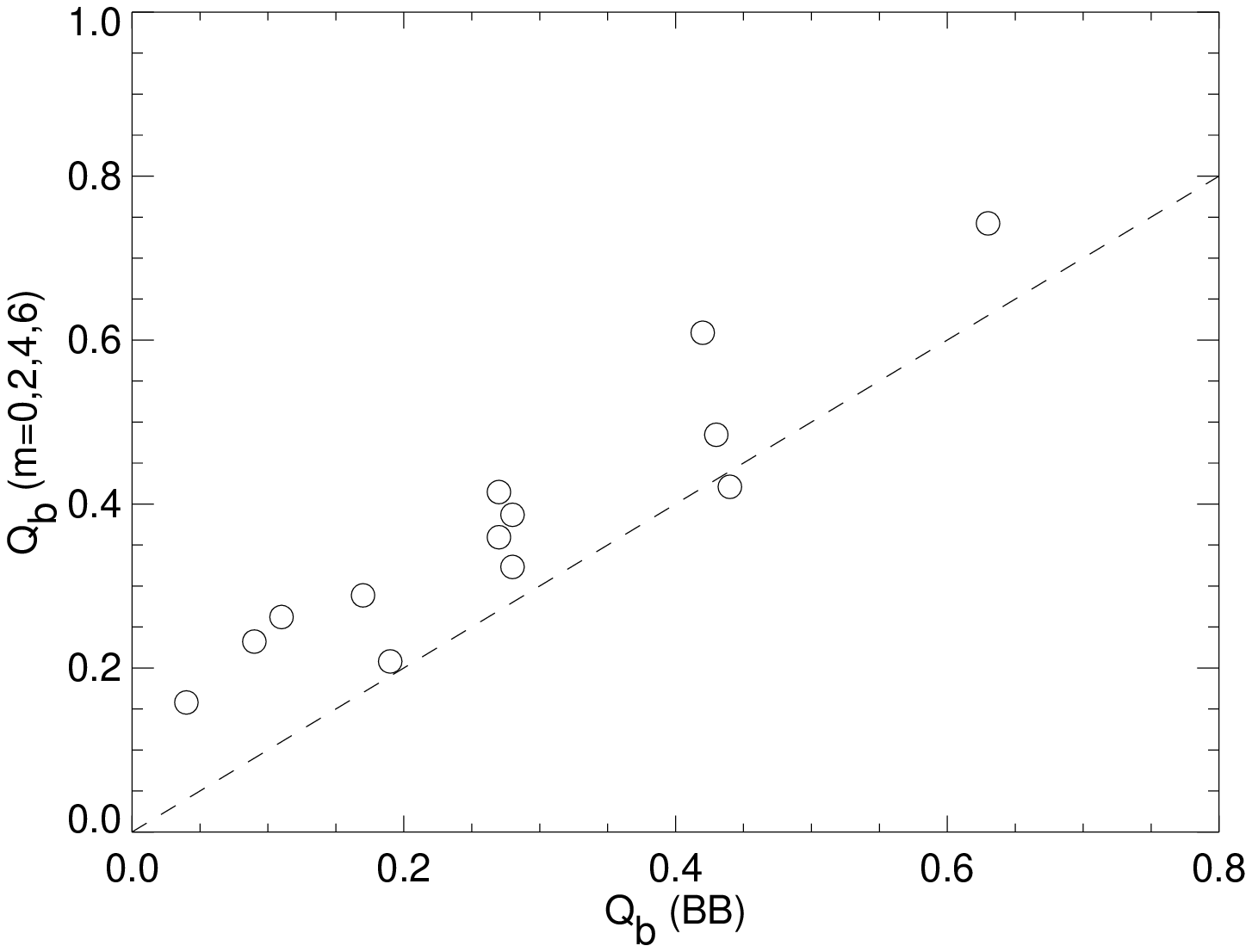,width=16cm} 
Fig. 6
\vfill
\eject

\psfig{file=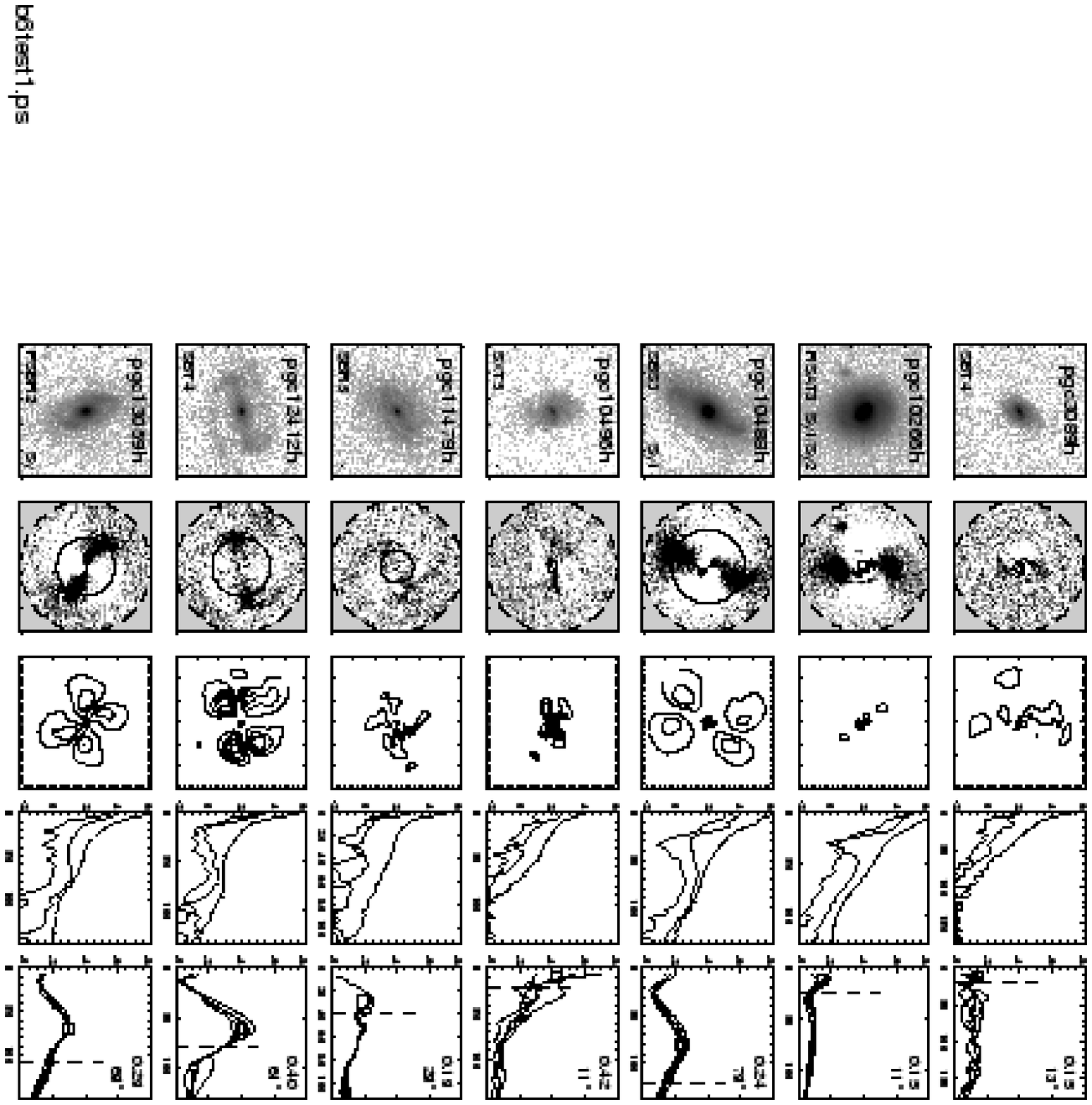,width=16cm} 
Fig. 7a
\vfill
\eject

\psfig{file=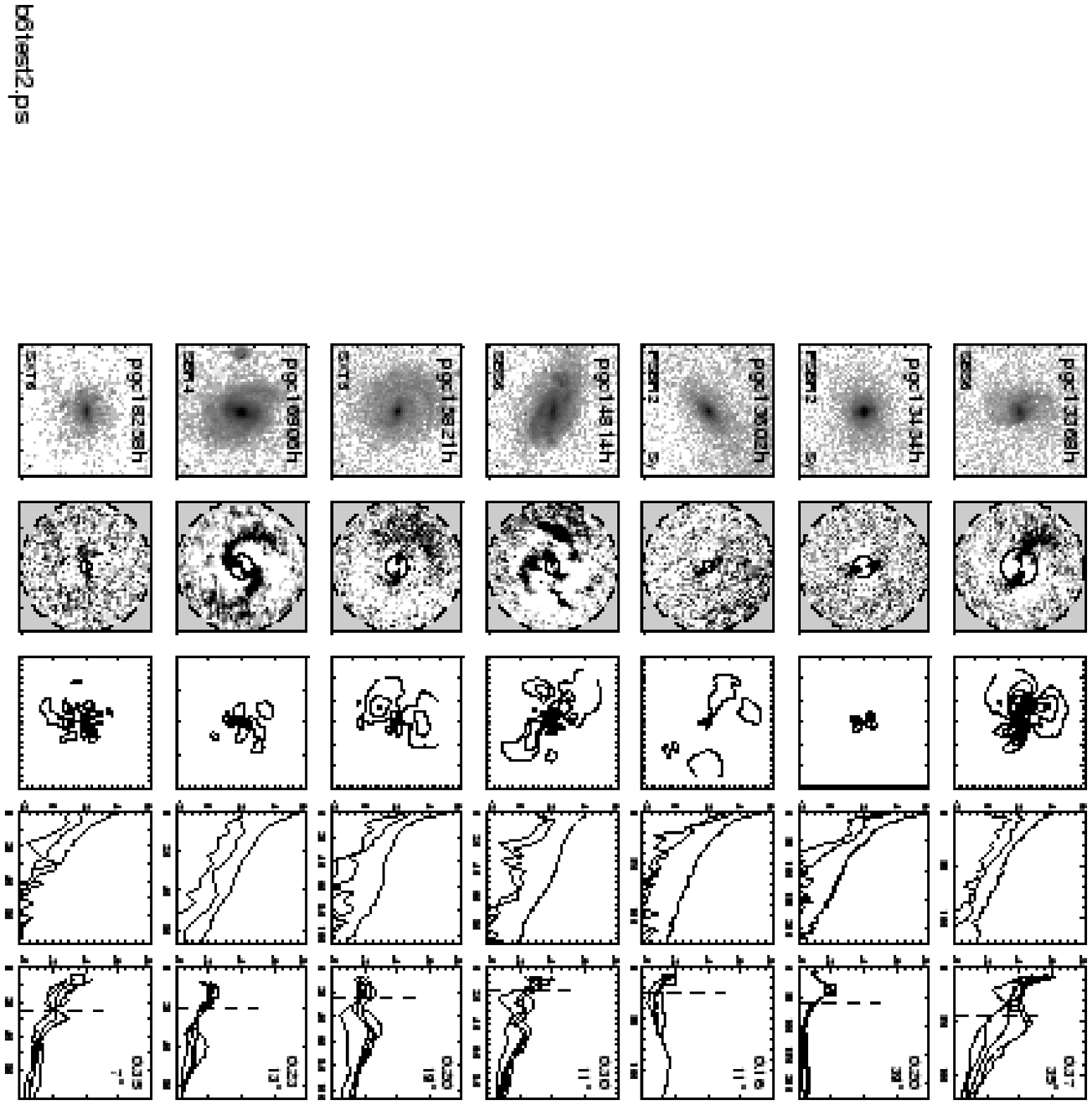,width=16cm} 
Fig. 7b
\vfill
\eject

\psfig{file=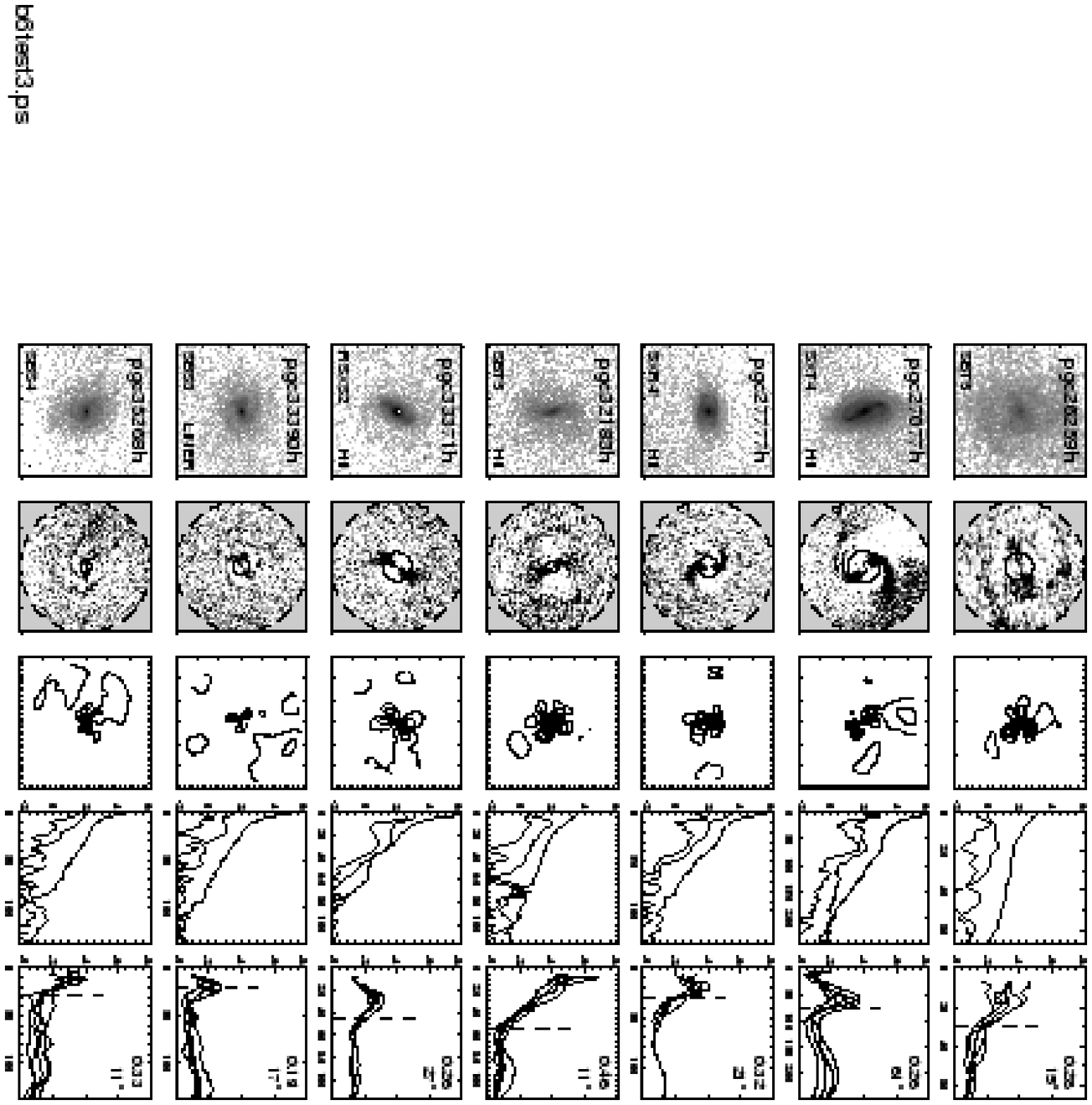,width=16cm} 
Fig. 7c
\vfill
\eject

\psfig{file=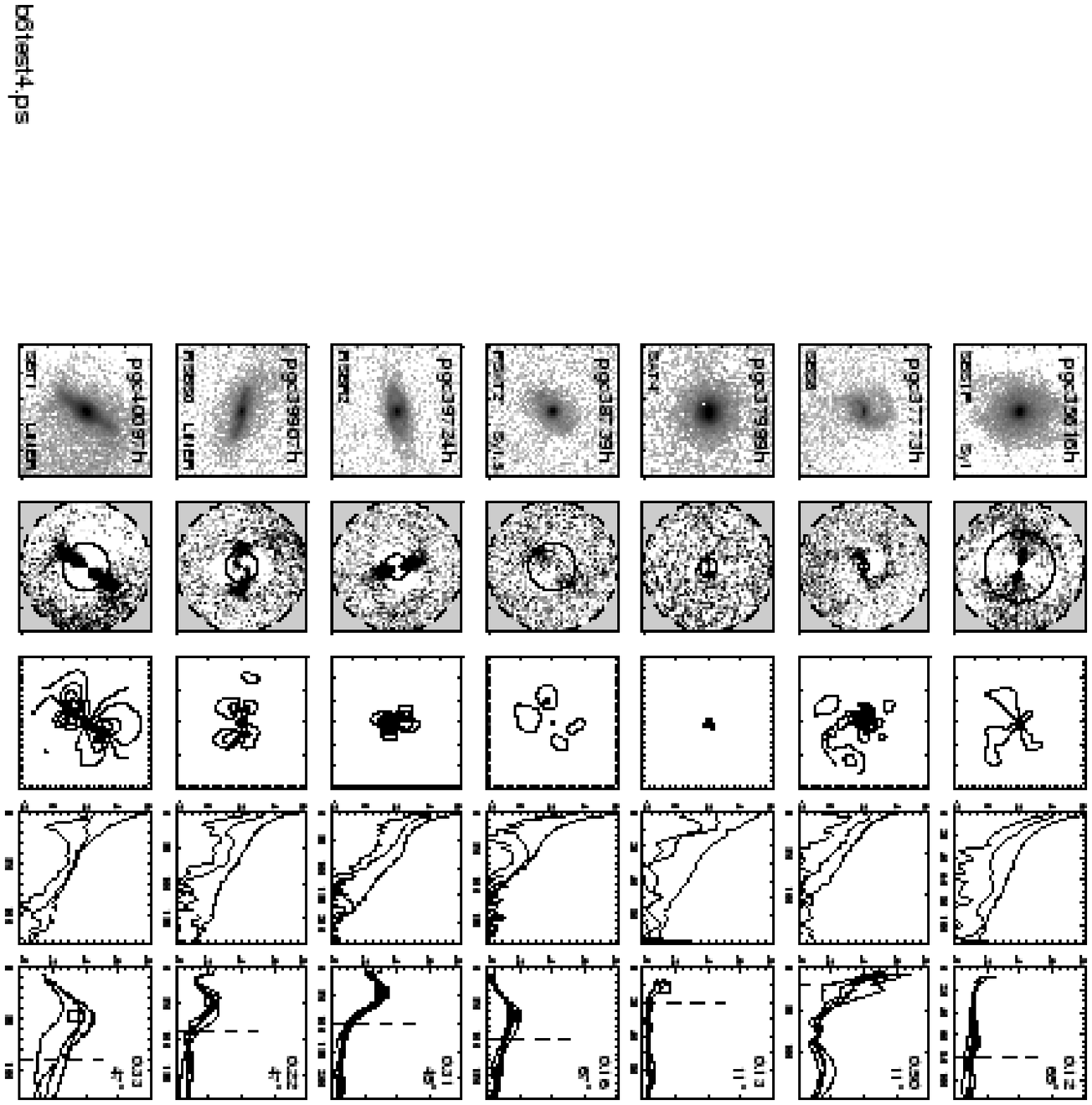,width=16cm} 
Fig. 7d
\vfill
\eject

\psfig{file=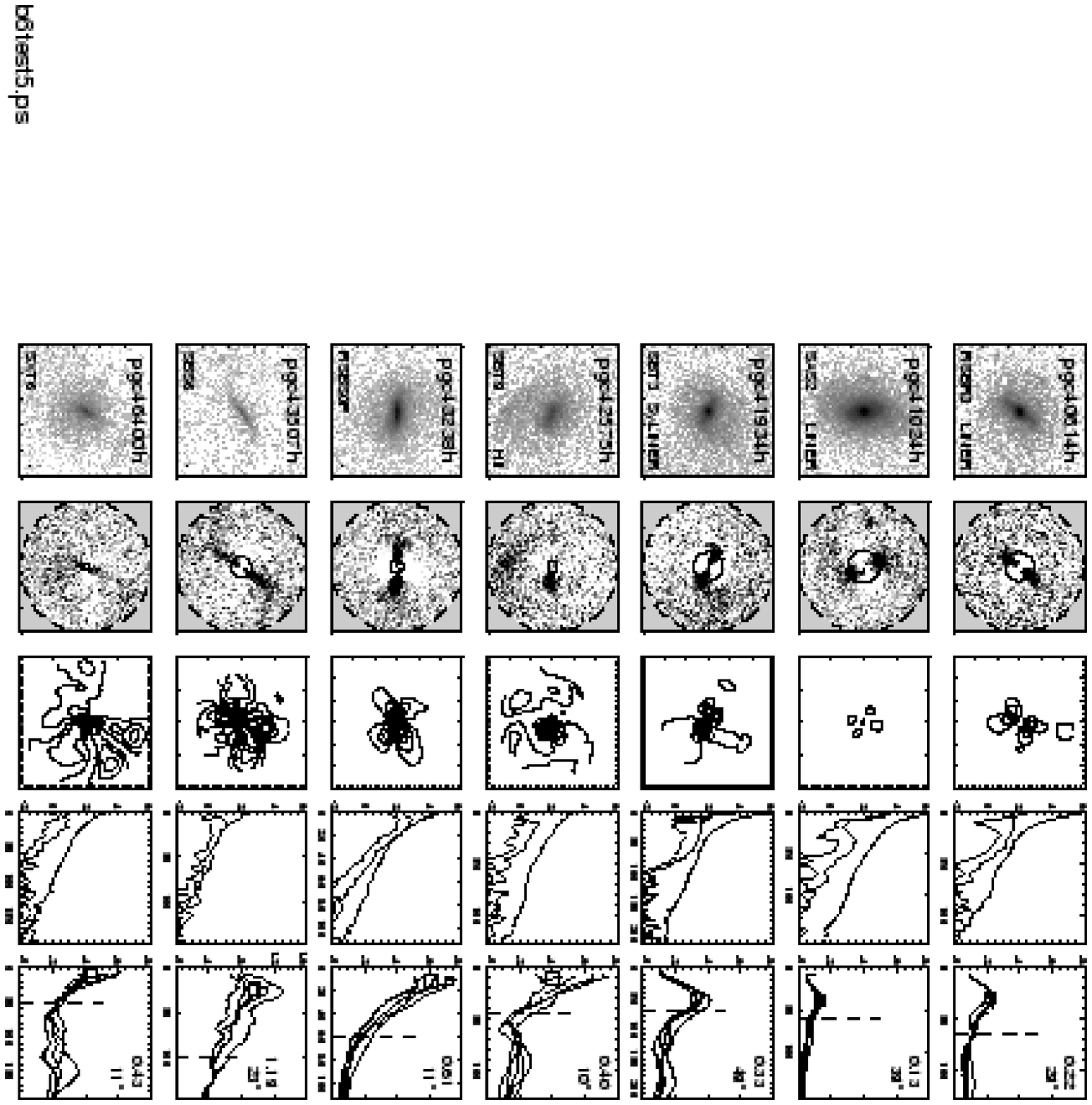,width=16cm} 
Fig. 7e
\vfill
\eject

\psfig{file=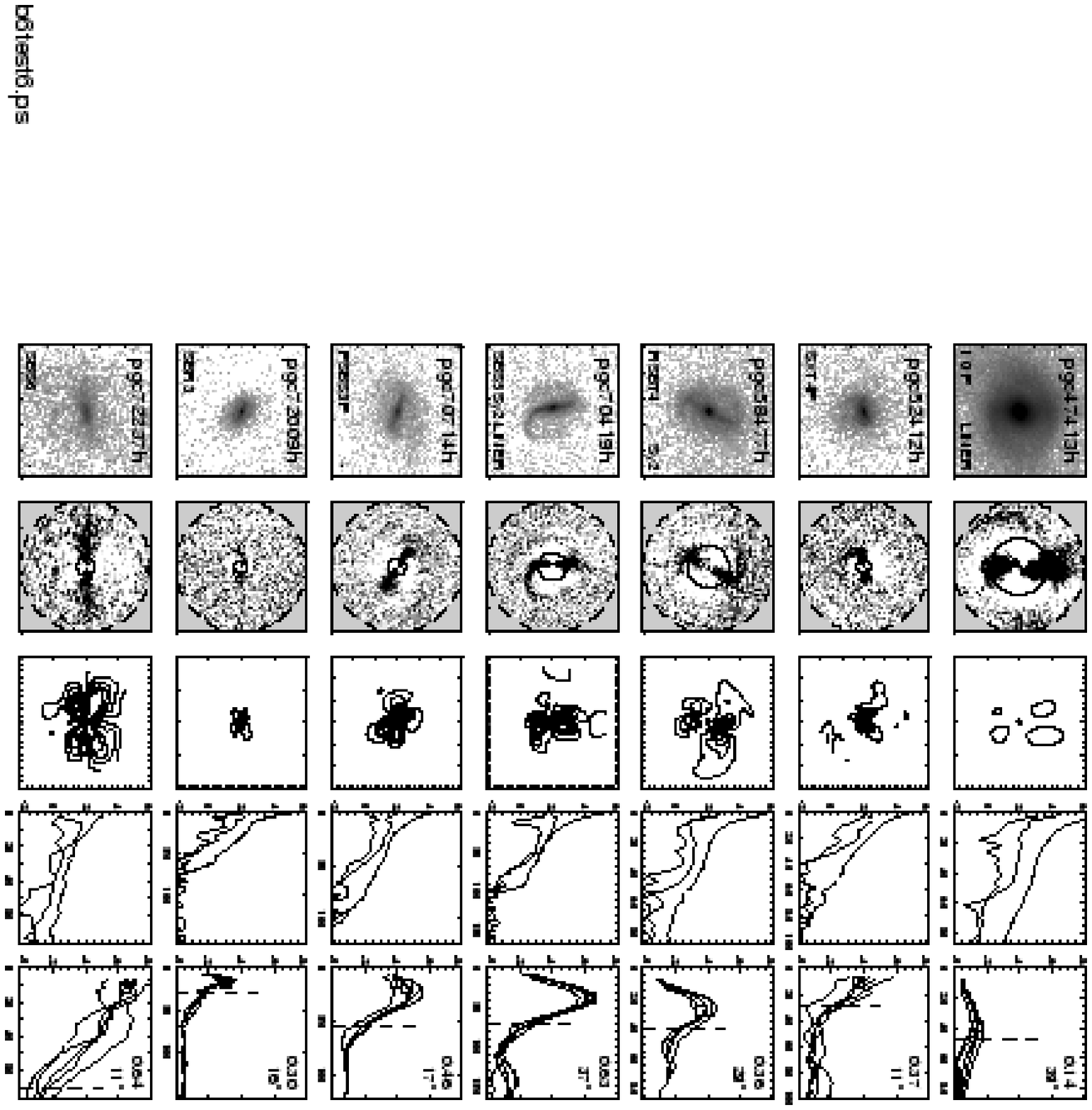,width=16cm} 
Fig. 7f
\vfill
\eject

\psfig{file=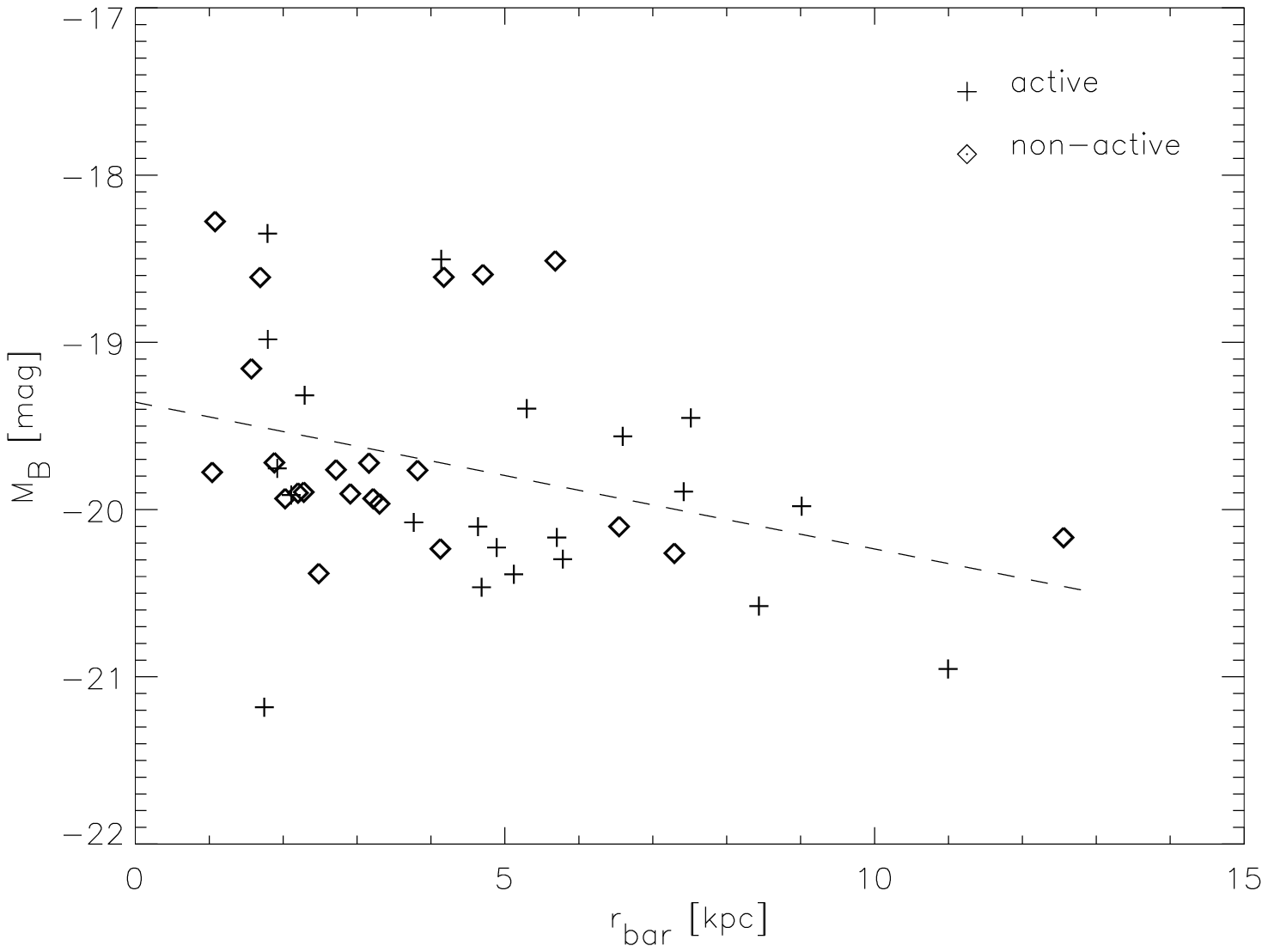,width=16cm} 
Fig. 8a
\vfill
\eject

\psfig{file=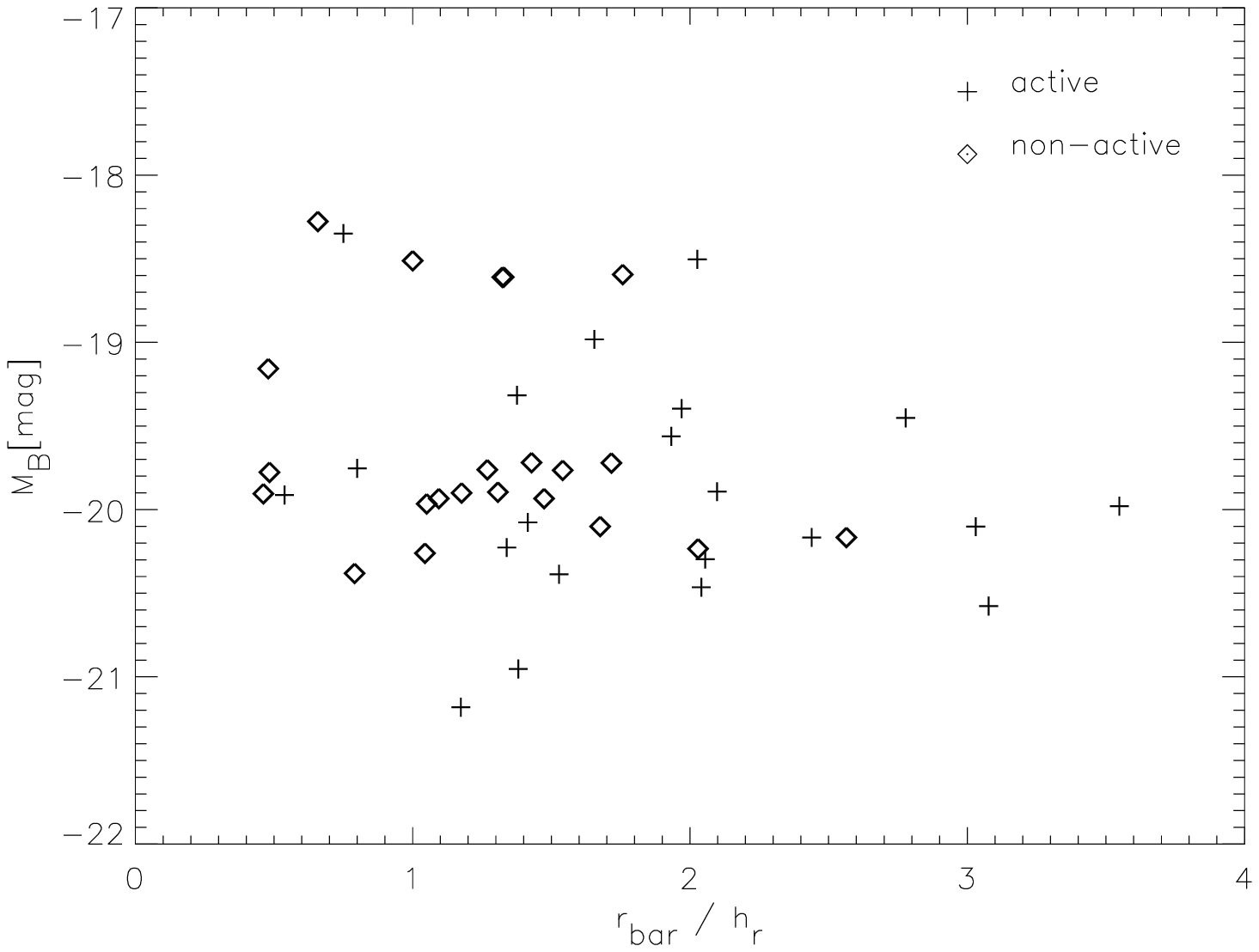,width=16cm} 
Fig. 8b
\vfill
\eject

\psfig{file=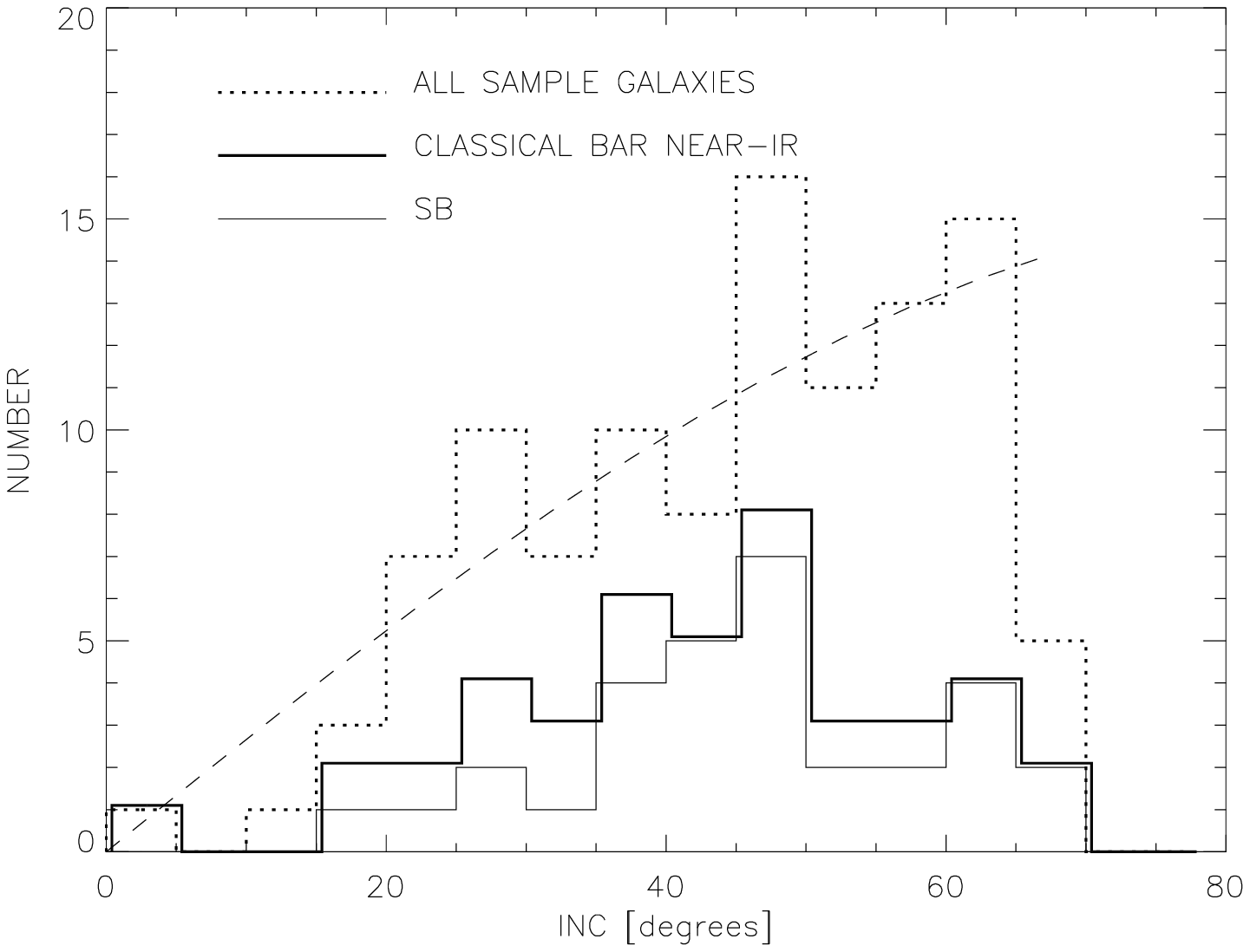,width=16cm} 
Fig. 9
\vfill
\eject

\psfig{file=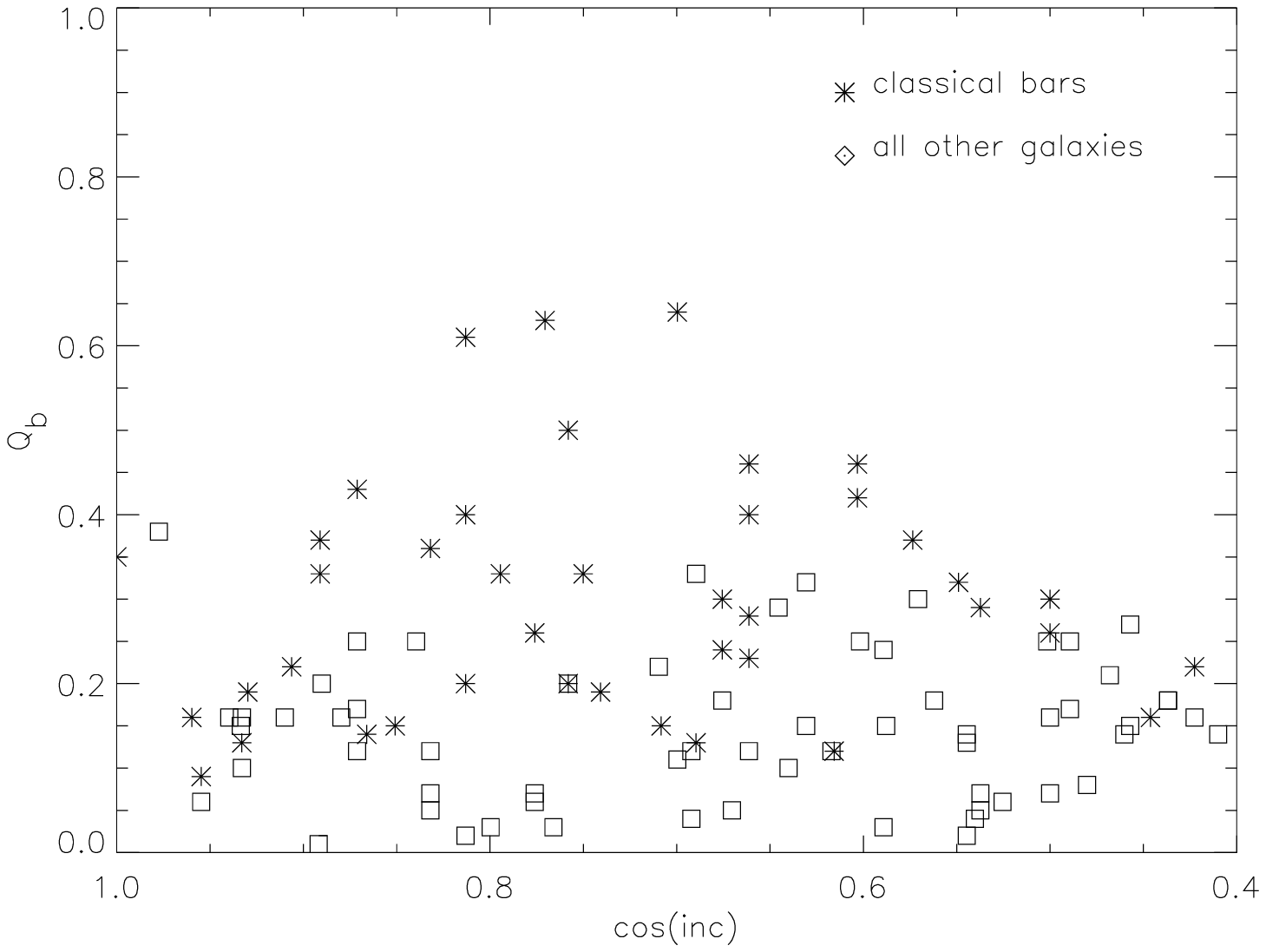,width=16cm} 
Fig. 10
\vfill
\eject

\psfig{file=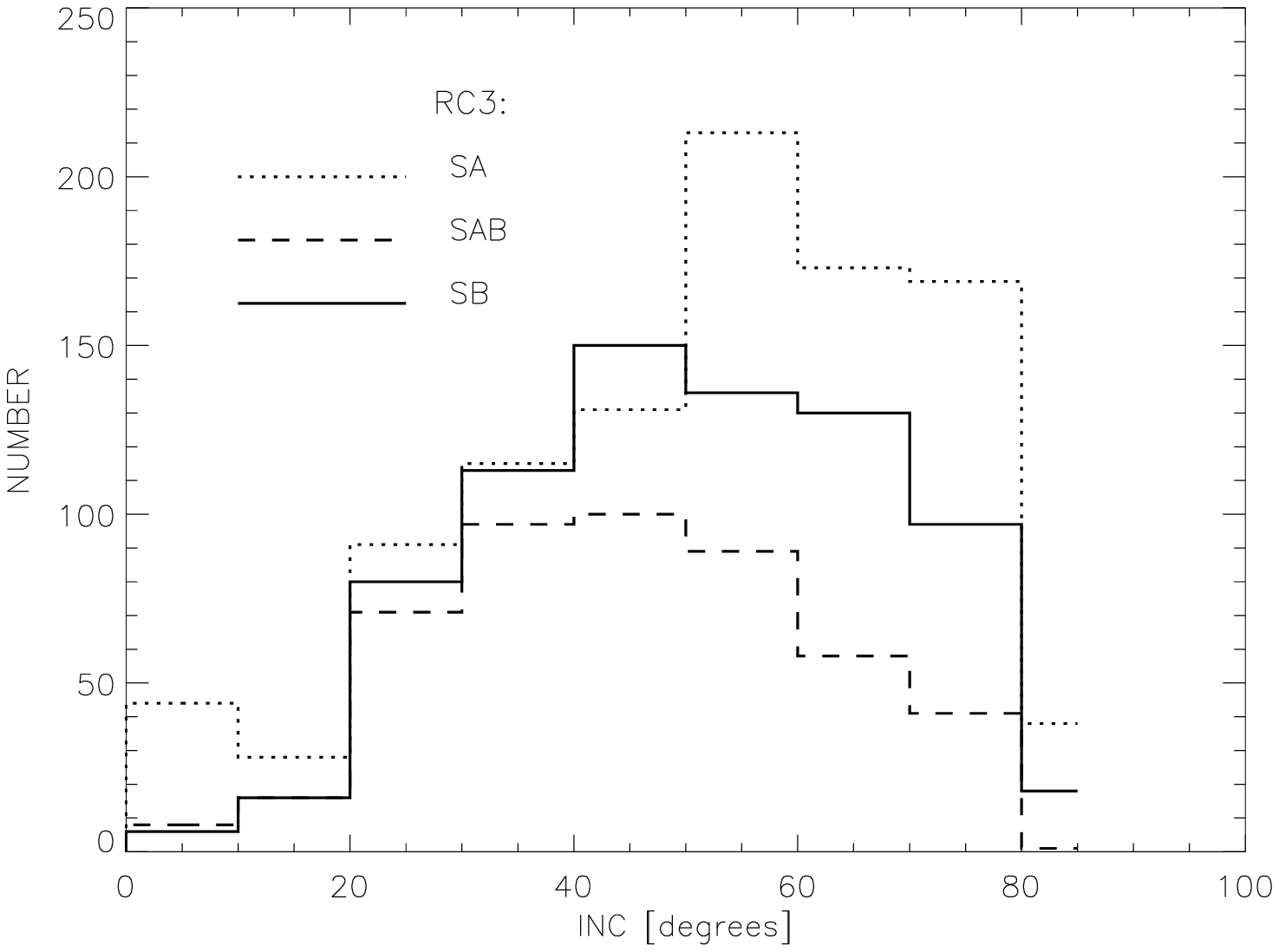,width=16cm} 
Fig. 11
\vfill
\eject

\psfig{file=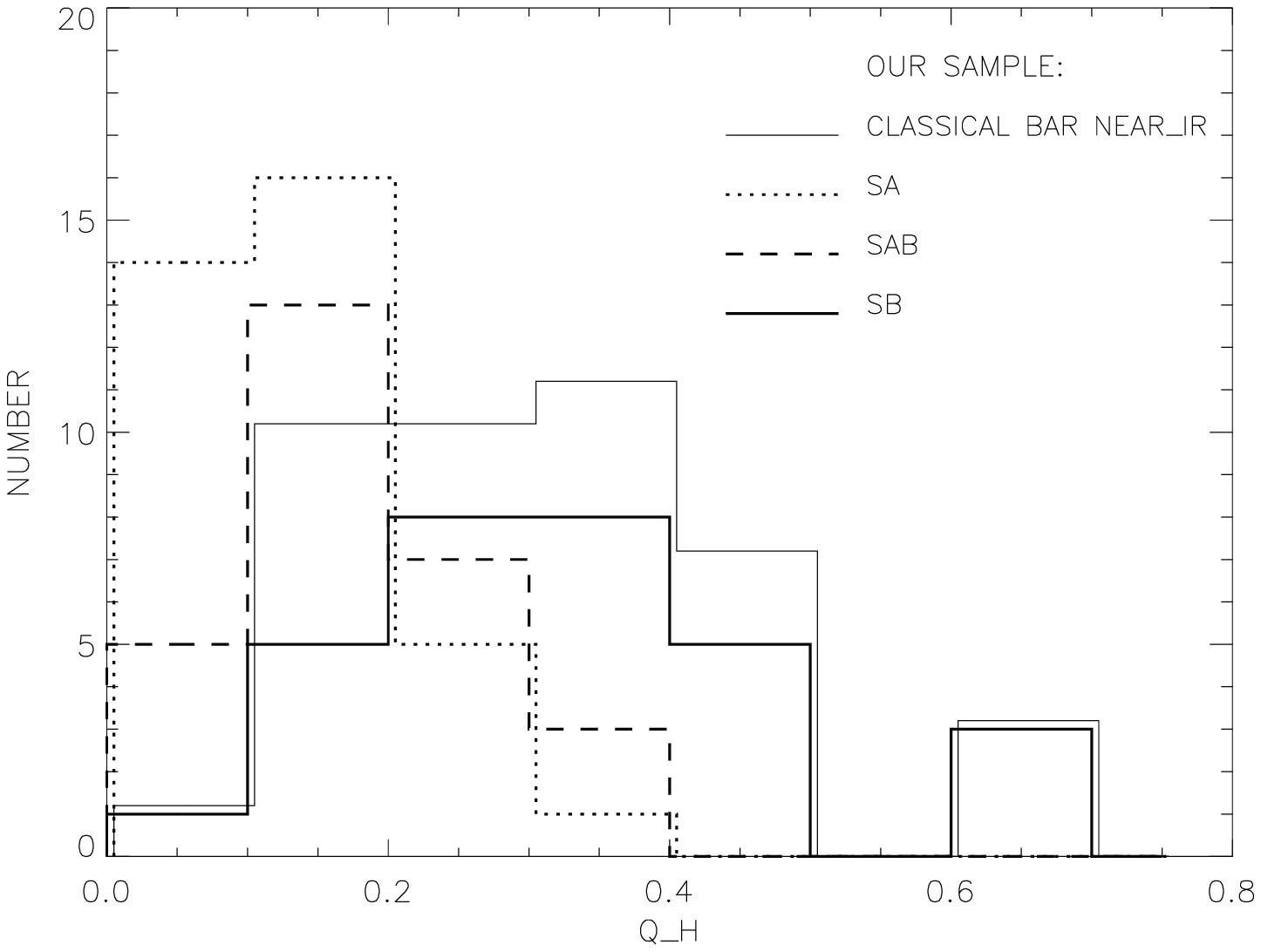,width=16cm} 
10Fig. 12a
\vfill
\eject

\psfig{file=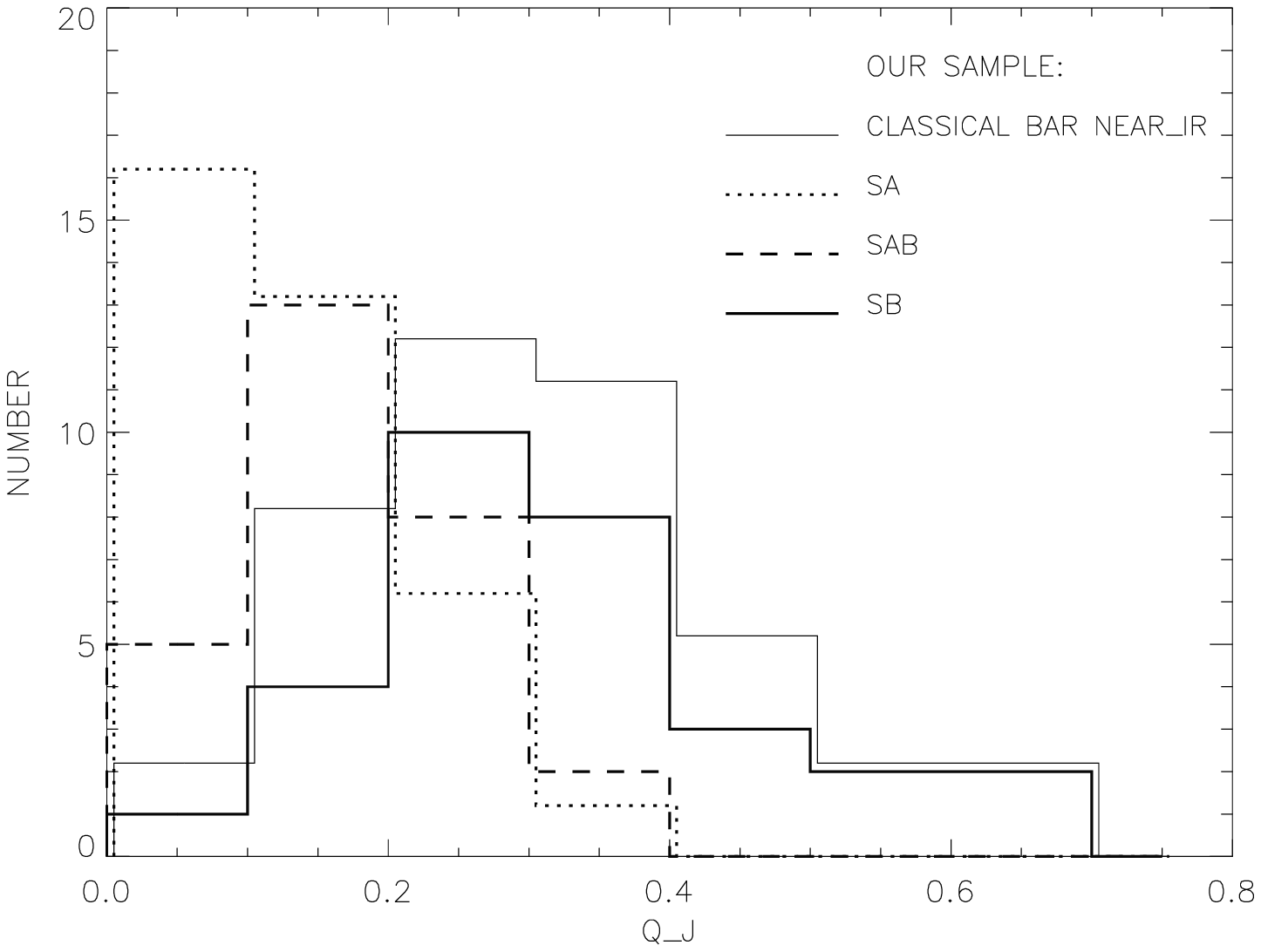,width=16cm} 
Fig. 12b
\vfill
\eject

\psfig{file=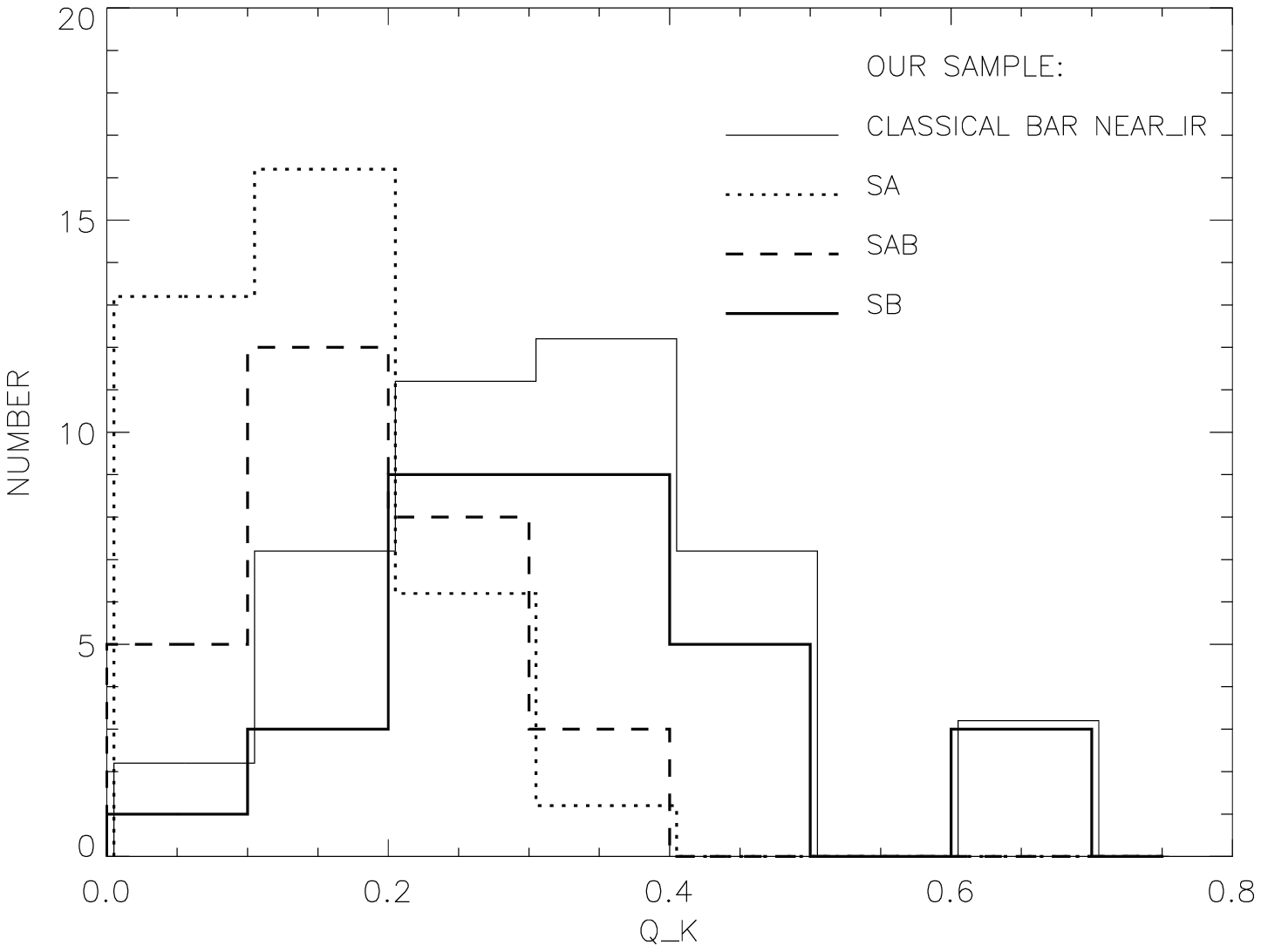,width=16cm} 
Fig. 12c
\vfill
\eject

\bye